\documentclass[12pt]{article}
\usepackage[margin=1in]{geometry}                
\usepackage{graphicx}
\usepackage{amssymb}
\usepackage{amsmath}
\usepackage{amsthm}

\usepackage{xcolor}
\usepackage{epstopdf}
 \usepackage{xr-hyper}
\usepackage{mathtools}
\usepackage[unicode=true,pdfusetitle,
 bookmarks=true,bookmarksnumbered=false,bookmarksopen=false,
 breaklinks=false,pdfborder={0 0 1},backref=false,colorlinks=true]
 {hyperref}
 \hypersetup{urlcolor=blue}
\hypersetup{citecolor=blue}
\hypersetup{filecolor=red}
\usepackage{url}
\usepackage{setspace}
\makeatletter
\def\@setthanks{\vspace{-\baselineskip}\def\thanks##1{\@par##1\@addpunct.}\thankses}
\makeatother
\usepackage{natbib}
\usepackage{caption}
\usepackage{tabularx}
 \usepackage{float}
 \usepackage{booktabs}
 \usepackage{multirow}
\usepackage{bm}
\newtheorem{thm}{Theorem}[section]

\newtheorem{lem}{Lemma}[section]
\newtheorem{cor}{Corollary}[section]
\newtheorem{assum}{Assumption}
\newtheorem{prop}{Proposition}[section]
\numberwithin{equation}{section}
\newtheorem{exm}{Example}[section]
\theoremstyle{remark}
\newtheorem{remark}{Remark}[section]
\newtheorem{innercustomgeneric}{\customgenericname}
\providecommand{\customgenericname}{}
\newcommand{\newcustomtheorem}[2]{%
  \newenvironment{#1}[1]
  {%
   \renewcommand\customgenericname{#2}%
   \renewcommand\theinnercustomgeneric{##1}%
   \innercustomgeneric
  }
  {\endinnercustomgeneric}
}
\numberwithin{equation}{section}
\newcustomtheorem{customthm}{Theorem}
\newcustomtheorem{customlem}{Lemma}
\newcustomtheorem{customassumption}{Assumption}
\newcustomtheorem{customprop}{Proposition}
\newcustomtheorem{customexample}{Example}
\newcustomtheorem{customdef}{Definition}
\newcustomtheorem{customcor}{Corollary}
\newcustomtheorem{customrem}{Remark}

\newtheorem{appxrem}{Remark}[section]
\usepackage{rotating}
\usepackage[title]{appendix}

\usepackage{todonotes}

\begin{document}
\begin{titlepage}

\title{Individual Welfare Analysis: Random Quasilinear Utility, Independence, and Confidence Bounds\thanks{We are grateful for the helpful comments provided by the Editor, an anonymous Associate Editor, and two anonymous referees. We have been benefited from comments from Songnian Chen, Lihua Lei, Yuya Sasaki, Benjamin Scuderi, Jia Xiang, and from seminar participants at HKUST, the University of Macau, the Peking University, the University of Queensland, CUHK-Shenzhen, the Econometric Society 2023 North American Summer Meeting and the 17th International Symposium on Econometric Theory and Applications. Feng gratefully acknowledges financial support from the Hong Kong Research Grants Council under Early Career Scheme Grant 26501721.}}
\author{Junlong Feng\thanks{Department of Economics, the Hong Kong University of Science and Technology. Email: \href{mailto:jlfeng@ust.hk}{jlfeng@ust.hk}.}\ \ \ \ \ \  Sokbae Lee\thanks{Department of Economics, Columbia University. Email: \href{mailto:sl3841@columbia.edu}{sl3841@columbia.edu}.}}

\date{November 2024}                                          

\maketitle
\vspace{-3em}
\begin{abstract} 
We introduce a novel framework for individual-level welfare analysis. It builds on a parametric model for continuous demand with a quasilinear utility function, allowing for heterogeneous coefficients and unobserved individual-good-level preference shocks. We obtain bounds on the individual-level consumer welfare loss at any confidence level due to a hypothetical price increase, solving a scalable optimization problem constrained by a novel confidence set under an independence restriction. This confidence set is computationally simple and robust to weak instruments, nonlinearity, and partial identification. The validity of the confidence set is guaranteed by our new results on the joint limiting distribution of the independence test by \cite{chatterjee2020}. These results together with the confidence set may have applications beyond welfare analysis. Monte Carlo simulations and two empirical applications on gasoline and food demand demonstrate the effectiveness of our method.\\
\vspace{0in}
\vspace{0in}
\end{abstract}
\end{titlepage}

\section{Introduction}

Consumer welfare analysis is an important topic in microeconomics. However, commonly adopted methods for welfare analysis are typically at the aggregated or averaged level, unable to recover individual heterogeneity. This paper proposes a novel framework to conduct individual-level welfare inference under a simple demand model.
More specifically, we aim to obtain confidence bounds on the welfare loss under a hypothetical price increase for an individual whose consumption level before the price hike is pre-specified.

Our framework builds on a parametric model for continuous demand following \cite{brown2007nonparametric} and \cite{BDW}. The model has a quasilinear utility function that allows for multiple goods and unobservables. Structural parameters in the model can depend on observables such as income or can be individual-specific. We derive the welfare loss under a price increase as a function of the parameters. Suppose we have a $(1-\alpha)\in (0,1)$ confidence set for these parameters. We can then infer the maximum and minimum of the welfare loss with confidence $(1-\alpha)$ by solving an optimization problem constrained by the confidence set. We show that under the chosen utility function, the welfare loss function is strictly increasing in each parameter, so the welfare loss bounds are very easy to obtain. Moreover, the function is also concave in the parameters. Therefore, if a researcher wishes to impose additional convex constraints on the parameters, the maximal welfare loss bound can still be quickly computed  even for each individual in the sample with a large number of goods.

Individual-level welfare bounds provide a deeper understanding of the data as they allow for the examination of heterogeneity within the data. Unlike a point counterfactual estimate, which is commonly produced through econometric procedures in structural modeling, our bounds are designed to incorporate a specified level of confidence. This means that there is no need for any additional computationally intensive inferential procedures to be conducted at the individual level.

To implement the method, we need first to obtain a confidence set at the desired confidence level for the structural parameters. We propose a new box confidence set under independence. The set is constructed by inverting a recently developed test of independence between two random variables by \cite{chatterjee2020}. Our confidence set is robust to nonlinearity, weak instruments, and partial identification. It can have other applications in different areas of economics beyond welfare analysis.

To guarantee the desired joint coverage probability of our confidence set for the multidimensional parameters on multiple goods, we prove the following result. For two independent continuous random vectors $\{Z_{k}\}_{k=1}^K$ and $\{W_{k}\}_{k=1}^K$ for some $K>1$, the asymptotic distribution of the vector of Chatterjee's statistics evaluated at $(W_{k},Z_{k})$ for each $k$ is jointly normal with variances all equal to 0.4 and covariances all equal to 0 under a mild condition. Notably, we allow for almost arbitrary correlation among $W_{k}$s and also among $Z_{k}$s across $k$; in particular, we allow some of or all the $Z_{k}$s to be identical. Under such asymptotic independence between these statistics, a multidimensional confidence set can be easily constructed by building multiple one-dimensional confidence intervals, enhancing computation efficiency. The new asymptotic result is general as it does not rely on the demand model we consider. We believe it is of independent interest.

To obtain tight bounds on welfare loss, it is desirable to have a tight confidence set as the constraint. Our proposed confidence set may not always be bounded, similar to the inversion of the Anderson-Rubin statistic. To overcome this issue, we can intersect our confidence set with an alternative confidence set obtained through a different method. In our demand model, we can construct a confidence set for the parameters by 2SLS or OLS with nonlinear transformations. Due to the nonlinear nature of the model, this alternative confidence set may also be large. The intersection of this set and the confidence set by inverting Chatterjee's test can be tighter than either of them, as observed in our empirical applications even with a small sample size.

To control the coverage of the intersection of the two confidence sets, we need to choose the coverage probability for each set appropriately. For this purpose, we derive the joint asymptotic distribution of a vector of Chatterjee's statistics under independence and a large class of asymptotically linear estimators based on which the alternative confidence set is constructed. We show that Chatterjee's statistics are asymptotically independent of the estimators, so joint inference can be again easily carried out.

This paper speaks to two strands of literature, demand estimation with welfare calculation and inference under independence. There is a large literature on demand estimation. See \cite{dube2019microeconometric} for a review of recent developments. The typical approach constructs a plug-in estimator of welfare change using a point estimator of the parameters of a structural model. However, usually, there is no corresponding confidence interval for the welfare analysis. 

For inference under independence, one popular approach is \textit{estimation based}. That is, first estimate the parameters by exploiting the relationship between the joint and marginal distributions of the unobservable and the exogenous variable that is independent of it, derive the limiting distribution of the estimator, and construct a confidence set based on that. Examples of this approach include \cite{manski1983closest}, \cite{brown2002weighted}, \cite{BDW}, \cite{komunjer2010semi}, \cite{poirier2017efficient}
 and \cite{torgovitsky2017minimum}. Like any estimation-based inferential approach, the quality of the confidence set typically depends on the quality of the asymptotic approximation of the estimator. Hence, weak instruments (if instrumental variables are needed) or partial identification may complicate the inferential procedure or result in an invalid confidence set.

This paper adopts an alternative approach by inverting a test. Confidence sets obtained by inverting a test, such as the Anderson-Rubin confidence set, are more robust to weak instruments and partial identification. In principle, one can invert any test of independence to obtain a confidence set following the general idea of this paper. Examples of such tests include Hoeffding's $D$ \citep{hoeffding1948non}, Blum-Kiefer-Rosenblatt's $R$ \citep{blum1961distribution}, and Bergsma-Dassios-Yanagimoto's $\tau^{*}$ \citep{bergsma2014consistent,yanagimoto1970measures}. As documented in \cite{shi2022power}, these tests of independence have better power properties than Chatterjee's test we adopt. We choose Chatterjee’s test for two major reasons. First, it does not require bootstrap/simulation to obtain the critical value. One even does not need to estimate the asymptotic covariance matrix to obtain the critical value since we show that the limiting distribution of a vector of Chatterjee's statistics is jointly normal with known covariance matrix as mentioned before. In contrast, the critical values of all the other aforementioned tests need to be simulated/bootstrapped, adding computation complexity. Specifically, for the scalar case, the limiting distributions of those test statistics under independence are infinite weighted sums of demeaned $\chi_{1}^{2}$ random variables \citep{shi2022power}; the joint distribution for a vector of such statistics, to the best of our knowledge, is unknown. Second, inverting Chatterjee's test computes much faster than inverting the others. Although all these test statistics including Chatterjee's can be computed in $O(n\log n)$ time in theory, we find in our Monte Carlo simulations that our confidence set by inverting Chatterjee's test computes around 1,000 times faster than inverting the other tests.

The rest of the paper is organized as follows. Section \ref{sec.welfare} sets up the demand model and develops our framework of inferring the bounds on the welfare loss with confidence. Section \ref{sec.CI} proposes a new method to construct confidence sets under independence and states the new asymptotic results. Section \ref{sec.simulation} presents Monte Carlo simulations. Section \ref{sec.application} demonstrates two empirical examples of gasoline and food demand. The Appendices provide heuristics on the shape and size of our confidence set, implementation details, some extensions, robustness checks for our empirical applications, and all the proofs.

\section{Individual-Level Welfare Confidence Bounds}\label{sec.welfare}

We consider the demand model in \cite{BDW} (henceforth BDW for simplicity). Suppose there are $K>0$ goods of interest. For $k=1,\ldots,K$, let $Y_{k}$ and $P_{k}$ be the consumption and price of good $k$, respectively. Let $Y_{0}$ be the consumption of the num\'eraire good with $P_{0}=1$. Let $W_{k}$ be an unobserved nonnegative random variable that affects the consumer's utility for good $k$. Let the $K\times 1$ vectors $\bm{Y}, \bm{P}$ and $\bm{W}$ collect $Y_{k}$, $P_{k}$ and $W_{k}$ for $k=1,\ldots,K$ respectively. For a consumer with income level $I$, she maximizes her quasilinear utility $U(Y_{0},\bm{Y},\bm{W},\theta)$ subject to the budget constraint:
\begin{align}
\max_{Y_{0},\bm{Y}}\ \  U(Y_{0},\bm{Y},\bm{W},\bm{\theta})\coloneqq & Y_{0}+U_{0}\left(\bm{Y};\bm{\theta}\right)+\bm{W}'\bm{Y} \notag\\
=&Y_{0}+\sum_{k=1}^{K}\theta_{k}g_{k}(Y_{k})+\bm{W}'\bm{Y} \label{eq.utility}\\
s.t.\ \ Y_{0}+\bm{P}' \bm{Y}=I,\label{eq.budget}
\end{align}
where $\bm{\theta}\coloneqq (\theta_{k})\in (0,1)^{K}$. The function $g_{k}$ is a known concave and strictly increasing function for all $k$. 

The model \eqref{eq.utility} captures good-specific consumer heterogeneity in the sense that it allows for an unobserved preference shock $W_k$ for each good $k$. For concreteness, let $g_{k}(\cdot) = \log(\cdot)$ for each $k$. This choice of $U_{0}$, together with $Y_{0}$ and $\bm{W}'\bm{Y}$, makes $U$ the logarithm of some \textit{perturbed} Cobb-Douglas utility function. We also allow $U_{0}$ to take other functional forms, or in principle allow $\bm{W}$ to be nonseparable. We provide more details on these extensions in Appendix \ref{sec:generalization_of_the_utility_function} and Remark \ref{rem.pum}, respectively.

On the other hand, we maintain linearity in $Y_{0}$ even in those extensions. Such linearity makes the utility we consider a special case of a random quasilinear utility function
proposed by \cite{brown2007nonparametric} that develops
theoretical results for quasilinear rationalizations of consumer demand data. Under a quasilinear utility function,  welfare changes due to a shift in price can be equivalently measured by the change in consumer surplus, compensating variation (CV), or equivalent variation (EV): all the three measures coincide and are equal to the difference in the indirect utility functions at the new and old prices. This property is useful to gain computational efficiency for the empirical method we propose.

One major limitation of quasilinearity utility is the absence of the income effect. Similar to the idea of \cite{griffith2018income} that allows income to enter the utility function in a flexible way to capture income effects in a discrete choice model, we mitigate this issue of missing income effect by allowing $\bm{\theta}$ to depend on income groups. We consider such a model in our first empirical application in Section \ref{sec.gas}.  Alternatively, when a panel data set is available, we can allow $\bm{\theta}$ (and more fundamentally $g_{k}$s) to differ across individuals but constant across time. In this case, the individual-wise income effect can exist; we consider this model in the empirical application in Section \ref{sec.food}.

We now derive the welfare loss function. Assuming an interior solution for the optimal $Y_{0}$ and $\bm{Y}$, BDW derives the following first-order condition for the optimal consumption $\bm{Y}^{*}(\bm{P},\bm{W};\bm{\theta})$: 
\begin{equation}
W_k = P_k - \frac{\theta_k}{Y_{k}^{*}(\bm{P},\bm{W};\bm{\theta})}, \; k =  1,\ldots, K.\label{eq.foc}
\end{equation}
 The model exploits separability, so that the first order condition for each good $k$ only involves $(W_k, P_k, \theta_k, Y^{*}_k)$ but not those from other goods.   
It follows from \eqref{eq.foc} that the demand for good $k$ is 
\begin{align}\label{eq.demand}
Y_k^{*}(\bm{P},\bm{W};\bm{\theta}) = \frac{\theta_k}{P_k - W_k}.
\end{align}
Substituting the budget constraint \eqref{eq.budget} and the optimal demand \eqref{eq.demand}, we then obtain the following indirect utility function from \eqref{eq.utility}:
\begin{align}\label{eq.iuf}
V(I, \bm{P}, \bm{W}, \bm{\theta}) := I + \sum_{k=1}^K \theta_k \log \frac{ \theta_k} { (P_k - W_k) } - \sum_{k=1}^K \theta_k.
\end{align}

We now focus on a change in consumer welfare due to a change in prices: the difference in the indirect utility function at the old and new prices by quasilinearity. Specifically,
let $\bar{\bm{w}}$ be a vector of realized $\bm{W}$ and ($\bm{p}^{0}$, $\bm{p}^{1}$) be two price levels; neither $\bm{p}^{0}$ nor $\bm{p}^{1}$ needs to be the actual observed price. Denote the optimal consumption of goods $1,\ldots,k$ under $\bm{p}^{j}$ and $\bar{\bm{w}}$ by $\bm{y}^{j}$ and the optimal consumption of the num\'eraire by $y_{0}^{j}$; by equation \eqref{eq.demand}, $\bm{y}^{j}\coloneqq \bm{Y}^{*}(\bm{p}^{j},\bar{\bm{w}};\bm{\theta})$, $j=0,1$. Assume that $I$ is sufficiently large and $\bar{\bm{w}}$ is smaller than both $\bm{p}^{0}$ and $\bm{p}^{1}$ so that the optimal consumption levels are interior solutions. These are automatically satisfied if $y_{0}^{0},\bm{y}^{0}>\bm{0}$ and $\bm{p}^{1}\geq \bm{p}^{0}$; see Remark \ref{rem.negative} for details. 
Let $\bm{\Delta} :=  \bm{p}^{1} - \bm{p}^{0}$ be the vector of price changes. By the definition of $\bm{y}^{0}$ and $\bm{y}^{1}$, equation \eqref{eq.iuf} implies that the welfare loss denoted by $\text{WL}(\bm{\theta})$, when the price changes from $\bm{p}^{0}$ to $\bm{p}^{1}$ while $\bm{W}$ is fixed at $\bar{\bm{w}}$, is as follows:
\begin{align}
\mathrm{WL}(\bm{\theta})&\coloneqq V(I, \bm{p}^{0}, \bar{\bm{w}}, \bm{\theta}) - V(I, \bm{p}^{1}, \bar{\bm{w}}, \bm{\theta})\notag\\
&= \sum_{k=1}^K \theta_k \log \left[ \frac{p_{k}^{1} - \bar{w}_{k}} {p_{k}^{0} - \bar{w}_{k}} \right] \notag\\
&= \sum_{k=1}^K \theta_k \log \left[ 1 + \frac{ \Delta_{k} y_{k}^{0}} { \theta_{k} } \right].\label{eq.CS}
\end{align}
Hence, the welfare loss solely depends on $\bm{y}^{0}$, the price change $\bm{\Delta}$, and the unknown parameters $\bm{\theta}$. 

So far, all the prices in equations \eqref{eq.foc}-\eqref{eq.CS} are real terms under the normalization that the price of the num\'eraire is 1. In most applications, however, we can only observe nominal prices in data. Denoting the nominal price for good $k=0,\ldots,K$ by $P_{k}^{n}$, we have $P^{n}_{k}/P_{k}\equiv P^{n}_{0}/1$ for every $k$. Equation \eqref{eq.foc} written in terms of the nominal prices then becomes $P_{k}^{n}-P_{0}^{n}W_{k}=P^{n}_{0}\theta_{k}/Y_{k}$.
If we only observe $P_{k}^{n}$ and $Y_{k}$ in data, we can in fact only treat $W_{k}^{n}\coloneqq P_{0}^{n}W_{k}$ as the unobservable and $\theta^{n}_{k}\coloneqq P^{n}_{0}\theta_{k}$ as the parameter. Consequently, although the real $\theta_{k}$ is assumed to lie in $(0,1)$ by BDW, we do not restrict the confidence interval to be upper bounded by 1 in the empirical applications: The parameter that data can recover is $\bm{\theta}^{n}$, which is $P^{n}_{0}$ times of the real $\bm{\theta}$ and $P^{n}_{0}$ is unknown. 

As a result, if we do not observe the real prices in a data set, the welfare loss function can be at most identified up to scale. To see it, plugging $\bm{\theta}^{n}$ and the observable prices $\bm{P}^{n}$ into the welfare loss function by treating them as the actual parameters $\bm{\theta}$ and the real prices $\bm{P}$, we obtain $\text{WL}^{n}(\bm{\theta}^{n})$ which is proportional to the actual welfare loss:
\begin{align}
  \text{WL}^{n}(\bm{\theta}^{n})\coloneqq &\sum_{k=1}^{K}\theta_{k}^{n}\log\left[1+\frac{\Delta_{k}^{n}y^{0}_{k}}{\theta_{k}^{n}}\right]\notag\\
 = &\sum_{k=1}^{K}(\theta_{k}P_{0}^{n})\log\left[1+\frac{(\Delta_{k}P_{0}^{n})y^{0}_{k}}{(\theta_{k}P_{0}^{n})}\right]\notag\\
  =&P_{0}^{n}\sum_{k=1}^{K}\theta_{k}\log\left[1+\frac{\Delta_{k}y^{0}_{k}}{\theta_{k}}\right]\notag\\
  =&P_{0}^{n}\text{WL}(\bm{\theta}),\label{eq.real}
\end{align}
where $\Delta_{k}^{n}$ is the nominal price change. Hence, the welfare loss that we can make inferences about based on nominal price data is $P_{0}^{n}$ times the actual welfare loss.

Nonetheless, we can still identify the following standardized welfare loss function based on nominal data:
\begin{align*}
\text{WL}^{st}(\bm{\theta})\coloneqq \frac{\text{WL}(\bm{\theta})}{\|\bm{\Delta}\|}=\frac{P_{0}^{n}\text{WL}(\bm{\theta})}{\|\bm{\Delta}P_{0}^{n}\|}\overset{(1)}{=}\frac{\text{WL}^{n}(\bm{\theta}^{n})}{\|\bm{\Delta}^{n}\|}\eqqcolon \text{WL}^{n,st}(\bm{\theta}^{n}),
\end{align*}
where $\|\cdot\|$ is the Euclidean norm of a vector. Equality (1) is by equation \eqref{eq.real}. The standardized welfare loss $\text{WL}^{st}(\bm{\theta})$ measures the welfare loss given a unit price change. The above equation shows that the standardized welfare loss is invariant to how the prices are measured. In the rest of the paper, we focus on finding confidence bounds on the standardized welfare loss. Since we can identify it with nominal prices, with a bit of abuse of notation, we still denote the nominal price $P^{n}_{k}$ and the parameter $\theta_{k}^{n}$ by $P_{k}$ and $\theta_{k}$ respectively for simplicity.

\begin{remark}\label{rem.corner}
The model in general can have corner solutions: Optimal consumption of the num\'eraire good is zero if $W_{k}\geq P_{k}$ for some $k$ or $I\leq \sum_{k}(P_{k}\theta_{k}/(P_{k}-W_{k}))$. In that case, the model is effectively no longer quasilinear and our welfare loss function is invalid. For good of interest $k=1,\ldots,K$, on the other hand, if the derivative of $g_{k}(\cdot)$ at 0, denoted by $g'_{k}(0)$, is well-defined, then zero consumption occurs when $W_{k}+g_{k}'(0)\leq P_{k}$. In this case, since we cannot write down a reduced form that expresses $\bm{W}$ as a function of the observables and parameters, we cannot form moment equations for the parameters. We focus on interior solutions throughout the paper and leave the study of corner solutions to future research. Hence, for simplicity, we pick $g_{k}(\cdot)=\log(\cdot)$ and assume $W_{k}<P_{k}$ for all $k$ with sufficiently large $I$ such that $I> \sum_{k}(P_{k}\theta_{k}/(P_{k}-W_{k}))$, then all the optimal $Y_{0}^{*},Y_{1}^{*},\ldots,Y_{K}^{*}$ are strictly positive.
\end{remark}

\begin{remark}\label{rem.pum}
Our perturbed Cobb-Douglas utility function is related to the perturbed utility model (PUM) in \cite{allen2019identification} with several key differences. In our notation, PUM assumes a more restrictive $U_{0}(\bm{Y};\bm{\theta})=\sum_{k}Y_{k}\theta_{k}$ whereas our $U_{0}$ can be nonlinear in $Y_{k}$s. On the other hand, our $\bm{W}'\bm{Y}$ is replaced in their work by a more general nonparametric function $D(\bm{Y},\bm{W})$. Their $\theta_{k}$s are functions of regressors; identification of $\theta_{k}$s requires that for every $k$, there exists a continuous $k$-specific regressor only entering $\theta_{k}$. In contrast, our approach does not rely on \textit{continuous} variation in the regressors: in our baseline model, the $\theta_{k}$s do not depend on any regressor, whereas in Section \ref{sec.gas}, they only depend on some discrete variation of a regressor. Another benefit of adopting the linear form for $\bm{W}$ for our purpose is that the welfare loss function does not depend on the realization of $\bm{W}$, simplifying the analysis. To accommodate a model that is nonseparable in $\bm{W}$ is possible. In that case, the welfare loss function in general depends on the realization of $\bm{W}$. One can either set $\bm{W}$ to be a counterfactual value or use its actual realization in a data set if that can be backed out using the observed consumption and price.
\end{remark}

\subsection{Predicting the Individual Welfare Loss Bounds with Confidence} 
 Suppose that we have micro-level i.i.d. data on $\{(P_{k,i}, Y_{k,i}, {\bm{X}_{i}}):i=1,\ldots,n;k=1,\ldots,K\}$, where subscript $i$ refers to observation $i$. 
Vector $\bm{X}_{i}$ contains other observables in the data. For a given price change $\bm{\Delta}$ {and consumption level $\bm{y}^{0}$}, it remains to deal with the unknown $\bm{\theta}$ {to compute the welfare change}. {In principle, we allow $\bm{\theta}$ to depend on $i$ as long as there are sufficient data to make inference about it; each $\theta_{k}$ can be either a nonparametric function of the other observables $\bm{X}_{i}$, or contain $i$-specific fixed coefficients; in that case, we allow $\bm{y}^{0}$ and $\bm{\Delta}$ to be $i$-specific as well. In our empirical applications, the varying coefficient model in Section \ref{sec.gas} and the model in Section \ref{sec.food} follow these two setups respectively. For notational simplicity, we treat $\bm{\theta}$, $\bm{\Delta}$ and $\bm{y}^{0}$ as homogeneous across $i$ in this section and in Section \ref{sec.CI}.}

For a given $\alpha\in (0,1)$, assume that we have a $(1-\alpha)$ confidence set for $\bm{\theta}$. We can then predict the maximal and minimal standardized welfare loss with confidence $(1-\alpha)$ for an individual with consumption {$\bm{y}^{0}\equiv\{y^{0}_{1},\ldots,y^{0}_{K}\}$} due to a price change $\bm{\Delta} \equiv (\Delta_1,\ldots,\Delta_K)$. Denoting the confidence set by $CS(1-\alpha)$, we solve the following problem: 
\begin{align}
\max_{\tilde{\bm{\theta}}} \text{(or $\min_{\tilde{\bm{\theta}}}$)}&\ \  \frac{1}{\sum_{k=1}^{K}\Delta_{k}^{2}}\sum_{k=1}^K \tilde{\theta}_k \log \left[ 1 + \frac{ \Delta_{k} } { \tilde{\theta}_k } {y_{k}^{0}} \right],\label{eq.objst}\\
s.t.&\ \  \tilde{\bm{\theta}}\in CS(1-\alpha).\notag
\end{align}
\begin{thm}\label{thm.welfare}
Let the maximum and the minimum of the objective function under the constraint above be $\overline{\mathrm{WL}}^{st}(1-\alpha)$ and $\underline{\mathrm{WL}}^{st}(1-\alpha)$, respectively. For each $k=1,\ldots,K$, if $\Pr(\bm{\theta}\in CS(1-\alpha))\to (1-\alpha)$ with the sample size $n\to \infty$, then 
\begin{align*}
\lim_{n\to\infty}\Pr\left(\mathrm{WL}^{st}(\bm{\theta})\leq\overline{\mathrm{WL}}^{st}(1-\alpha)\right)\geq& (1-\alpha),\\
\lim_{n\to\infty}\Pr\left(\mathrm{WL}^{st}(\bm{\theta})\geq\underline{\mathrm{WL}}^{st}(1-\alpha)\right)\geq&(1-\alpha),\\\
\lim_{n\to\infty}\Pr\left(\underline{\mathrm{WL}}^{st}(1-\alpha)\leq\mathrm{WL}^{st}(\bm{\theta})\leq\overline{\mathrm{WL}}^{st}(1-\alpha)\right)\geq& (1-\alpha).
\end{align*}
\end{thm}

In the next section, we will propose a method to construct a box, or, a rectangular confidence set $CS(1-\alpha)$. The optimization problem will then be very convenient to solve. As shown in Appendix \ref{appx.monotone}, the welfare loss $\text{WL}(\tilde{\bm{\theta}})$ as a function of $\tilde{\bm{\theta}}\in\mathbb{R}^{K++}$ is strictly increasing in each $\tilde{\theta}_{k}$ on $(\max\{0,-\Delta_{k}y^{0}_{k}\},\infty)$ for all $\Delta_{k}y_{k}^{0}\neq 0$ and $\Delta_{k}y_{k}^{0}>-\theta_{k}$. Hence, for instance, under a price increase such that all $\Delta_{k}$s are positive, if $CS(1-\alpha)$ is rectangular as the cartesian product of confidence intervals of the $\theta_{k}$s, one can easily obtain the maximum and minimum of the welfare loss by substituting the upper and lower bounds of those intervals, respectively. When there are additional convex constraints besides the convex confidence set, for instance, $\sum_{k=1}^{K}\tilde{\theta}_{k}=1$, one can still solve the maximization problem quickly by off-the-shelf convex optimization algorithms because the objective function is concave in $\bm{\theta}$; see Appendix \ref{appx.implement} for more details. 

Individual-level welfare bounds contain more information than a bound at the aggregated level; one can view and study the heterogeneity from them. Moreover, our bound is different from a point estimate; by construction, it incorporates the desired confidence level, so no further inferential procedure is needed.

\begin{remark}\label{rem.negative}
We only consider a price increase in this paper, i.e., $\Delta_{k}\geq 0$ for all $k>0$ with the inequality being strict for some $k$. This is a simple sufficient (but not necessary) condition to make sure that if $y_{0}^{0}>0$ and $\bm{y}^{0}>\bm{0}$ satisfies equation \eqref{eq.demand}, meaning that the optimal consumption including the num\'eraire good is an interior solution at some income level, then $\bm{y}^{1}$ also satisfies equation \eqref{eq.demand}. Consequently, the welfare loss function we derive is valid. To see this, $\bm{y}^{0}$ satisfying equation \eqref{eq.demand} implies that the realization of the unobservable $\bar{\bm{w}}$ satisfies $\bm{p}^{0}>\bar{\bm{w}}$. Hence, if the new prices $\bm{p}^{1}>\bm{p}^{0}$, we have $\bm{p}^{1}>\bar{\bm{w}}$. Meanwhile, the total expenditure for the $K$ goods of interest $\sum_{k}(p_{k}^{1}\theta_{k}/(p_{k}^{1}-w_{k}))\leq \sum_{k}(p_{k}^{0}\theta_{k}/(p_{k}^{0}-w_{k}))$. So, the new optimal consumption including the num\'eraire good is an interior solution under the original budget. Nonetheless, we can analyze the welfare change under a price decrease, too, so long as the optimal consumption at the new price is still an interior solution. To satisfy the interior solution requirements, we first need $p_{k}^{1}-\bar{w}_{k}=\Delta_{k}+p_{k}^{0}-\bar{w}_{k}>0$, which leads to $\Delta_{k}>-\theta_{k}/y_{k}^{0}$; this is exactly the condition that makes the right-hand side of equation \eqref{eq.CS} well-defined. So, for a price drop ($\Delta_{k}<0$), if we have a box confidence set $CS(1-\alpha)$ formed by the cartesian product of confidence intervals for each $\theta_{k}$, then we can compute the welfare loss bounds under $\Delta_{k}$ as long as $\Delta_{k}>-\text{LB}_{k}/y^{0}_{k}$ where $\text{LB}_{k}>0$ is the lower bound of the confidence interval for $\theta_{k}$. In the meantime, we assume that the budget $I$ is sufficiently high, greater than the expenditure of the $K$ goods of interest under the new price.
 \end{remark}

\section{A New Confidence Set for Nonlinear Models}\label{sec.CI}

So far, we have assumed the existence of a box confidence set for $\bm{\theta}$. In this section, we introduce a novel approach to obtain it. This approach is based on a new correlation coefficient to measure independence developed by \cite{chatterjee2020}. We first briefly review this statistic for completeness in Section \ref{sec.chatterjee}. We then derive a new result about the joint limiting distribution of this coefficient for multiple pairs of independent random variables in Section \ref{sec.joint}; this result enables us to invert individual test statistics to construct a joint box confidence set for $\bm{\theta}$. Section \ref{sec.intersect} derives the joint limiting distribution of Chatterjee's correlation coefficients and a general class of estimators. Built on it, we propose an intersection method to sharpen the confidence set.

\subsection{Review of Chatterjee's Correlation Coefficient}\label{sec.chatterjee}

Consider an i.i.d. sample $\{(R^{(1)}_{i},R^{(2)}_{i}):i=1,\ldots,n\}$ for random variables $R^{(1)}$ and $R^{(2)}$. Suppose that the $R^{(1)}_{i}$s and $R^{(2)}_{i}$s have no ties.\footnote{When there are ties in the $R_{i}^{(1)}$s, \cite{chatterjee2020} proposes to break the ties uniformly at random. When there are ties in the $R^{(2)}_{i}$s, define $l_{i}$ as the number of $j$ such that $R^{(2)}_{(j)}\geq R^{(2)}_{(i)}$. Then redefine $\xi_{n}(R^{(1)},R^{(2)})\coloneqq 1-n\sum_{i=1}^{n-1}|r_{i+1}-r_{i}|/[2\sum_{i=1}^{n}l_{i}(n-l_{i})]$.} Sort the data by $R^{(1)}$ in ascending order. Let the rearranged data be $(R^{(1)}_{(1)},R^{(2)}_{(1)}),\ldots,(R^{(1)}_{(n)},R^{(2)}_{(n)})$ where $R^{(1)}_{(1)}<\cdots<R^{(1)}_{(n)}$. Let $r_{i}$ be the rank of $R^{(2)}_{(i)}$. \cite{chatterjee2020} proposes the following statistic to measure the dependence between $R^{(1)}$ and $R^{(2)}$: 
\begin{equation}
  \xi_{n}(R^{(1)},R^{(2)})\coloneqq 1-\frac{3\sum_{i=1}^{n-1}|r_{i+1}-r_{i}|}{n^{2}-1}.\label{eq.xi}
\end{equation}

When $R^{(1)}\perp R^{(2)}$, where $\perp$ denotes independence, Chatterjee shows in his Theorem 1.1 that $\xi_{n}(R^{(1)},R^{(2)})\overset{a.s.}{\to}0$. Meanwhile, $R^{(1)}\not\perp R^{(2)}$ if and only if $\text{plim } \xi_{n}(R^{(1)},R^{(2)})>0$. Moreover, under $R^{(1)}\perp R^{(2)}$, his Theorem 2.2 shows that 
\begin{equation*}
  \sqrt{n}\xi_{n}(R^{(1)},R^{(2)})\overset{d}{\to}\mathcal{N}(0,0.4),
\end{equation*}
 if $R^{(2)}$ is continuous. When $R^{(2)}$ is discrete, the asymptotic variance needs to be estimated and Chatterjee provides a consistent estimator for it which can be computed in time $O(n\log n)$.

\subsection{Joint Limiting Distribution}\label{sec.joint}
For $k=1,\ldots,K$, let $Z_{k}$ be an observed exogenous random variable for which we also have an i.i.d. sample. Depending on the application, variable $Z_{k}$ can be price itself or its instrumental variable. Let the short-hand notation $\xi_{n}^{(k)}$ denotes $\xi_{n}(W_{k},Z_{k})$. Let $\bm{\xi}_{n}\coloneqq (\xi_{n}^{(1)},\ldots,\xi_{n}^{(K)})$. Chatterjee establishes the limiting distribution of $\xi^{(k)}_{n}$ when $W_{k}\perp Z_{k}$. For joint inference of $\bm{\theta}$, we now derive the asymptotic distribution of $\bm{\xi}_{n}$.

\begin{assum}\label{assu.iid}
$\{(W_{1,i},\ldots,W_{K,i},Z_{1,i},\ldots,Z_{K,i}):i=1,\ldots,n\}$ are i.i.d.
\end{assum}
\begin{assum}\label{assu.cont}
i) Random vector $(W_{1},\ldots,W_{K},Z_{1},\ldots,Z_{K})$ is continuously distributed. ii) $(W_{1},\ldots,W_{K})\perp (Z_{1},\ldots,Z_{K})$.
\end{assum}
\begin{remark}\label{rem.Zequal}
We do not require there to be $K$ different $Z_{k}$s; the current setup is for convenience only. We allow that some of or even all these $Z_{k}$s are equal.
\end{remark}
The continuity requirement for the unobservables $W_{k}$ in Assumption \ref{assu.cont}-i) is common in economics. For $Z_{k}$, we only focus on continuous instruments in the applications of this paper. Asymptotic theory for the case of discrete instruments is left for future research. Under this assumption, there are no ties in a data set with probability one. For Assumption \ref{assu.cont}-ii),  although the independence assumption is strong, it is indispensable in many nonseparable models, which, in principle, we could handle as mentioned in Remark \ref{rem.pum}.

We now introduce an assumption regulating the dependence structure of the $W_{k}$s.
For $k=1,\ldots,K$, let $\pi_{k}:\{1,\ldots,n\}\mapsto\{1,\ldots,n\}$ be a random variable such that $\pi_{k}(i)$ indicates the location of the \textit{original} index $i$ in the permuted index set after arranging $W_{k,1},\ldots,W_{k,n}$ in ascending order. When there are no ties among $W_{k,i}$s, which has probability one under Assumption \ref{assu.cont}-i), $\pi_{k}(i)$ is the number of $W_{k,j}$s satisfying $W_{k,j}\leq W_{k,i}$. The mapping $\pi_{k}$ in this case is one-to-one. When there are ties, we can still uniquely define a one-to-one $\pi_{k}$ such that for any $W_{k,i}=W_{k,j}$ with $i\neq j$, we let $\pi_{k}(i)<\pi_{k}(j)$ if and only if $i<j$. 

Notation-wise, to distinguish the post-permutation indices under permutation $(\pi_{k}(1),\ldots,\pi_{k}(n))$ from the original ones, denote the post-permutation indices by $(i)_{k}$. In other words, $W_{k,(1)_{k}}\leq W_{k,(2)_{k}}\leq \cdots\leq W_{k,(n)_{k}}$. For example, suppose the $l$-th smallest element in $W_{k,i}$s (assuming there are no ties) is $W_{k,j}$, then $\pi_{k}(j)=(l)_{k}$. 
\begin{assum}\label{assum.dependence}
For any $k\neq k'$,
$$\Pr\left(\left|\pi_{k}(i)-\pi_{k}(j)\right|=1\cap \left|\pi_{k'}(i)-\pi_{k'}(j)\right|=1\right)=o\left(\frac{1}{n}\right),\forall 1\leq i<j\leq n.$$
\end{assum}
Assumption \ref{assum.dependence} says that for any original indices $i$ and $j$, if they are adjacent after arranging the $W_{k,i}$s in ascending order, then they are not likely to be adjacent after arranging the $W_{k',i}$s in ascending order. To be more precise, note that under Assumption \ref{assu.iid},
\begin{equation}\label{eq.2overn}
  \Pr\left(\left|\pi_{k}(i)-\pi_{k}(j)\right|=1\right)=\frac{2}{n},\forall i\neq j;
\end{equation}
see Appendix \ref{sec.proof.2overn} for a proof. Then, Assumption \ref{assum.dependence} is equivalent to $$\Pr\left(\left|\pi_{k'}(i)-\pi_{k'}(j)\right|=1\Big| \left|\pi_{k}(i)-\pi_{k}(j)\right|=1\right)=o\left(1\right),\forall 1\leq i<j\leq n.$$

Assumption \ref{assum.dependence} allows for correlated $W_{k}$s, but requires that $W_{k'}$ still has sufficient variation conditional on $W_{k}$. For further illustration, we now provide a sufficient condition for it, an example under which the sufficient condition holds, and a counterexample where Assumption \ref{assum.dependence} fails. Let $F_{W_{k}}$ denote the cumulative distribution function of $W_{k}$ for all $k$.

\begin{prop}\label{prop.suff}
Let $V_{k}\coloneqq F_{W_{k}}(W_{k})$ and $V_{k'}\coloneqq F_{W_{k}'}(W_{k}')$. If there exists some constant $C>0$ such that the following holds for all $v,v'\in (0,1)$:
\begin{equation}\label{eq.suff}
  \lim_{n\to\infty}\Pr\left(V_{k'}\in \left[v'-C\sqrt{\frac{\log\log{n}}{n}},v'+C\sqrt{\frac{\log\log{n}}{n}}\right]\Bigg|V_{k}=v \right)= 0,
\end{equation}
then Assumption \ref{assum.dependence} holds under Assumptions \ref{assu.iid} and Assumption \ref{assu.cont}-i).
\end{prop}
Condition \eqref{eq.suff} is mild. It says that conditional on any value of $V_{k}$, there is no atom in the distribution of $V_{k'}$. Note that Proposition \ref{prop.suff} does not rule out nonrectangle support for $(W_{k},W_{k'})$: For $v'$ such that $F^{-1}_{W_{k'}}(v')$ is no longer in the support of $W_{k'}|W_{k}=F_{W_{k}}^{-1}(v)$, condition \eqref{eq.suff} trivially holds.

We now provide an example and a counterexample.

\begin{exm}\label{exm.normal}
Condition \eqref{eq.suff} holds if $(W_{k},W_{k'})$ are jointly normal for any correlation coefficient $\rho_{k,k'}\in(-1,1)$. This is because conditional on any value of $W_{k}$, the random variable $W_{k'}$ is still continuously distributed on the entire real line as a normal distribution with variance equal to $(1-\rho_{k,k'}^{2})$ times its original variance. 
\end{exm}

\begin{exm}\label{exm.counter}
Suppose $W_{k}\sim \mathcal{N}(0,1)$. Let $D\sim Ber(0.5)$ be a dummy variable and $\widetilde{W}\sim \mathcal{N}(0,1)$. Assume $W_{k},D$ and $\widetilde{W}$ are mutually independent. Now suppose $W_{k'}=W_{k}D+\widetilde{W}(1-D)$. Then one can verify that for all $i\neq j$,
\begin{align*}
  &\Pr\left(\left|\pi_{k}(i)-\pi_{k}(j)\right|=1\cap \left|\pi_{k'}(i)-\pi_{k'}(j)\right|=1\right)
  \\
  =&\frac{1}{2}\Pr\left(\left|\pi_{k}(i)-\pi_{k}(j)\right|=1|D=1\right)+\frac{1}{2}\Pr\left(\left|\pi_{k}(i)-\pi_{k}(j)\right|=1\cap \left|\pi_{k'}(i)-\pi_{k'}(j)\right|=1|D=0\right)\\
  =&\frac{1}{2}\cdot\frac{2}{n}+\frac{1}{2}\cdot\left(\frac{2}{n}\right)^{2}
  \neq o\left(\frac{1}{n}\right),
\end{align*}
where the first equality is by the law of iterated expectation, by the distribution of $D$, and by $W_{k'}=W_{k}$ when $D=1$. The second equality is by mutual independence among $D,W_{k},\widetilde{W}$, by $W_{k'}=\widetilde{W}$ under $D=0$, and by equation \eqref{eq.2overn}. One can verify that condition \eqref{eq.suff} does not hold, too.
\end{exm}

These examples show that we can allow for arbitrarily high correlation among the unobservables, so long as there is no atom in the conditional distribution.

We are now ready to state our main result.
\begin{thm}\label{thm.joint1}
Under Assumptions \ref{assu.iid} to \ref{assum.dependence}, 
\begin{equation}
  \sqrt{\frac{n}{0.4}}\bm{\xi}_{n}\overset{d}{\to}\mathcal{N}(\bm{0},I_{K}),
\end{equation}
where $I_{K}$ is the $K\times K$ identity matrix.
\end{thm}
Theorem \ref{thm.joint1} says that under our assumptions, the $\sqrt{n}\xi^{(k)}_{n}$s are asymptotically jointly normal and mutually independent. 

Three remarks are in order. First, Assumption \ref{assum.dependence} is \textit{not needed} to obtain joint normality; it is a sufficient condition to obtain zero limiting covariances. So, for instance, if one believes that there is only one unobservable for all different goods, i.e., $W_{1}=\cdots = W_{K}$, then joint normality still holds but the asymptotic covariances may need to be estimated. Second, Assumption \ref{assum.dependence} is silent about the dependence structure of the instruments $Z_{k}$s. So Theorem \ref{thm.joint1} holds even if there is only one instrument; see also Remark \ref{rem.Zequal}.  Finally, Theorem \ref{thm.joint1} is independent of our demand model; it holds for any independent random pairs satisfying Assumptions \ref{assu.iid} to \ref{assum.dependence}.

Based on Theorem \ref{thm.joint1}, we propose the following procedure to construct a $(1-\alpha)$ joint confidence set for $\bm{\theta}$. Suppose there is a known parameter space $\Theta_{k}$ for each $\theta_{k}$. Let $z_{(1-\alpha)^{1/K}}$ denote the $(1-\alpha)^{1/K}$-th quantile of the standard normal distribution. For each $k$, construct
\begin{equation}\label{eq.CSindi}
  CS^{\xi}_{k;\alpha}\coloneqq\left\{\theta\in\Theta_{k}:\sqrt{\frac{n}{0.4}}\xi_{n}\left(P_{k}-\frac{\theta}{Y_{k}},Z_{k}\right)\leq z_{(1-\alpha)^{\frac{1}{K}}}\right\}
  \end{equation}
Then construct the joint confidence set by
\begin{equation}\label{eq.CSonly}
  CS(1-\alpha)=CS^{\xi}_{1;\alpha}\times CS^{\xi}_{2;\alpha}\times \cdots\times CS^{\xi}_{K;\alpha}.
\end{equation}

By Theorem \ref{thm.joint1}, the coverage probability for each $CS_{k;\alpha}^{\xi}$ is asymptotically $(1-\alpha)^{1/K}$ and the joint coverage probability is approaching $(1-\alpha)$. We formally state this result in the following corollary.
\begin{cor}\label{cor.cs}
Under Assumptions \ref{assu.iid}-\ref{assum.dependence}, for $CS(1-\alpha)$ constructed by equation \eqref{eq.CSonly},
\[
  \Pr\left(\bm{\theta}\in CS(1-\alpha) \right)\to 1-\alpha.
\]
 \end{cor}

Note that Corollary \ref{cor.cs} does not require any \textit{relevance condition}. The validity of the confidence set holds regardless of the strength of the instrument. Even if the model is only partially identified or the instrument is weak, $CS(1-\alpha)$ always yields the correct asymptotic coverage probability for $\bm{\theta}$. Moreover, one does not need to distinguish the point and partially identified cases. On the other hand, the strength of the instrument does affect the size of the confidence set. For instance, if $Z_{k}$ is independent of both $P_{k}$ and $W_{k}$, then one can see that $CS^{\xi}_{k;\alpha}$ can be the entire real line. Appendix \ref{appx.shape} provides more details on the shape and size of the confidence set.

In practice, one can do a grid search over $\Theta_{k}$ for each $k$ and take the minimum and maximum grid nodes that fall into the set \eqref{eq.CSindi} to get a confidence interval. The joint box confidence set can be constructed as the cartesian product of these intervals. Such a construction enables fast and scalable computation even when $K$ is large because we only need to construct $K$ individual confidence intervals by one-dimensional grid search.

\subsection{Intersecting Confidence Sets}\label{sec.intersect}

One possible drawback of Chatterjee's test of independence is its low power \citep{shi2022power}. As a consequence, the sets $CS^{\xi}_{k;\alpha}$s could be wide. In this section, we propose an intersection approach to mitigate the low power problem and tighten the confidence set.

For the demand model we consider, we can construct a confidence set for $\bm{\theta}$ based on the estimation approach instead of by inverting Chatterjee's $\xi$; that is, first estimate $\bm{\theta}$ and then build a confidence set based on the estimator. Specifically, rearranging equation \eqref{eq.foc}, we have $1/Y_{k}=\beta _{k}P_{k}-\beta_{k}W_{k}$ where $\beta_{k}=1/\theta_{k}$ for each $k$. Treating $1/Y_{k}$ as the dependent variable, we can estimate $\beta_{k}$ by OLS or seemingly unrelated regression (SUR) if we assume the prices are exogenous, or by 2SLS or multi-equation GMM if the prices are endogenous but we have instruments for them. One can then, for instance, obtain a sup-t confidence band for $\bm{\theta}$ with any desired coverage probability by the delta method and the procedure described in \cite{montiel2019simultaneous}. 

Now we can obtain a potentially tighter confidence set by intersecting the one introduced in Section \ref{sec.joint} with an estimation-based confidence set. To control the coverage probability, we derive the joint limiting distribution of $(\bm{\xi}_{n},\hat{\bm{\theta}})$ where estimator $\hat{\bm{\theta}}$ comes from a class of estimators specified in Assumption \ref{assum.expansion} below. Let $\widetilde{W}_{k,i}\coloneqq W_{k,i}-\mathbb{E}(W_{k,i})$ and $\widetilde{\bm{W}}_{i}\coloneqq(\widetilde{W}_{1,i},\ldots,\widetilde{W}_{K,i})'$. Let $\bm{Z}_{i}\coloneqq(Z_{1,i},\ldots,Z_{K,i})'$. We assume the estimator is asymptotically linear with the following influence function:  
\begin{assum}\label{assum.expansion}
There exist functions $\omega_{i}:\mathbb{R}^{K}\mapsto\mathbb{R}^{K\times K}$ such that
\begin{equation}
  \sqrt{n}\left(\hat{\bm{\theta}}-\bm{\theta}\right)=\frac{1}{\sqrt{n}}\sum_{i=1}^{n}\omega_{i}(\bm{Z}_{i})\widetilde{\bm{W}}_{i}+o_{p}(1),
\end{equation}
and
\begin{equation}
  \text{Var}\left(\frac{1}{\sqrt{n}}\sum_{i=1}^{n}\omega_{i}(\bm{Z}_{i})\widetilde{\bm{W}}_{i}\right)\overset{p}{\to}\Sigma,
\end{equation}
where the $K\times K$ matrix $\Sigma$ is positive definite and consistently estimable.
\end{assum}
One can verify that a large class of estimators including all the aforementioned estimators satisfy Assumption \ref{assum.expansion} under certain regularity conditions. The following Theorem \ref{thm.joint2} says that $\bm{\xi}_{n}$ and such an estimator $\hat{\bm{\theta}}$ are asymptotically independent under regularity conditions. Let $\omega_{(k,k'),i}$ be the $(k,k')$-th entry in matrix $\omega_{i}$.
\begin{thm}\label{thm.joint2}
Under Assumptions \ref{assu.iid}-\ref{assum.expansion}, if $\max_{k,k'}\mathbb{E}\left(\omega_{(k,k'),i}(\bm{Z}_{i})\widetilde{W}_{k',i}\right)^{8}<\infty$, then
\begin{equation}\label{eq.lim_dist_estimator}
  \sqrt{n}\begin{pmatrix}\frac{\bm{\xi}_{n}}{\sqrt{0.4}}\\ \hat{\bm{\theta}}-\bm{\theta}\end{pmatrix}\overset{d}{\to}\mathcal{N}\left(0,\begin{pmatrix}I_{K}&\bm{0}\\\bm{0}&\Sigma \end{pmatrix}\right).
\end{equation}
\end{thm}
 \begin{remark}
Asymptotic normality and independence between $\bm{\xi}_{n}$ and $\hat{\bm{\theta}}$ hold even if Assumption \ref{assum.dependence} is dropped. That is, without Assumption \ref{assum.dependence}, equation \eqref{eq.lim_dist_estimator} still holds by only replacing $I_{K}$ on the right-hand side with a possibly more complicated $K\times K$ matrix. 
 \end{remark}
 \begin{remark}
The existence of the eighth moment is required by the central limit theorem in \cite{chatterjee2008new} that we adopt; it is assumed only because of $\hat{\bm{\theta}}$.
 \end{remark}

Theorem \ref{thm.joint2} provides a simple way to construct a $(1-\alpha)$ confidence set for $\bm{\theta}$. Suppose we have a $\sqrt{1-\alpha}$ box confidence set for $\bm{\theta}$ based on Assumption \ref{assum.expansion}, formed as the cartesian product of individual intervals $\left[\hat{\theta}_{k}-c_{k;\alpha}/\sqrt{n},\hat{\theta}_{k}+c_{k;\alpha}/\sqrt{n}\right]$ where the positive constants $c_{k;\alpha}$s are known or can be consistently estimated. One way to obtain such a confidence set is by the sup-t method by \cite{montiel2019simultaneous}, which has exact $\sqrt{1-\alpha}$ asymptotic coverage. Then, one can construct an intersected confidence set $CS(1-\alpha)$ as follows: For each $k$, construct
\begin{equation}\label{eq.CSindiinter}
  CS^{inter}_{k;\alpha}\coloneqq\left\{\theta\in\left[\hat{\theta}_{k}-\frac{c_{k;\alpha}}{\sqrt{n}},\hat{\theta}_{k}+\frac{c_{k;\alpha}}{\sqrt{n}}\right]:\sqrt{\frac{n}{0.4}}\xi_{n}\left(P_{k}-\frac{\theta}{Y_{k}},Z_{k}\right)\leq z_{(1-\alpha)^{\frac{1}{2K}}}\right\}.
  \end{equation}
Then construct the joint confidence set by
\begin{equation}\label{eq.CSinter}
  CS(1-\alpha)=CS^{inter}_{1;\alpha}\times CS^{inter}_{2;\alpha}\times \cdots\times CS^{inter}_{K;\alpha}.
\end{equation}
\begin{cor}\label{cor.csinter}
Under Assumptions \ref{assu.iid}-\ref{assum.expansion}, for $CS(1-\alpha)$ constructed by equation \eqref{eq.CSinter},
 \[
  \Pr\left(\bm{\theta}\in CS(1-\alpha) \right)\to 1-\alpha.
\]
\end{cor}

Since the confidence sets by inverting Chatterjee's $\xi$ and by the estimation-based approach utilize information differently, it is possible that when using alone, neither is a subset of the other. Hence, the intersection approach may improve the power by yielding a smaller confidence set. In our Monte Carlo simulations and empirical applications, we do observe improvement in this intersection approach compared to using one method alone.

\begin{remark}
In principle, the estimation-based confidence set can be either an ellipsoid (e.g. Wald-type) or a box. Neither dominates the other \textit{per se} since they both can have exact $\sqrt{1-\alpha}$ asymptotic coverage probability. A box estimation-based confidence set, as we propose in this section, has the following advantages in our specific problem. First, it provides a compact parameter space for each $\theta_{k}$ so that a one-dimensional grid search is feasible to get $CS^{inter}_{k;\alpha}$ defined in equation \eqref{eq.CSindiinter}. Second, for the welfare loss optimization problem, although the maximum can still be easily solved for under an ellipsoid confidence set as a convex constraint, the minimum is difficult to obtain. In contrast, when the confidence set is a box, one can solve the constrained minimization problem simply by plugging in the lower bounds of the individual confidence intervals utilizing coordinate-wise strict monotonicity of the welfare loss function.
\end{remark}

\section{Monte Carlo Simulations}\label{sec.simulation}
\subsection{Comparison of Chatterjee's $\xi$ and Other Tests of Independence}\label{sec.MC.3}
In this section, in a model with three goods, we examine the performance of our confidence set by i) checking the rejection frequencies of the joint test, ii) examining the individual confidence intervals' length for each parameter and computation time and iii) computing the welfare loss bounds under various constraints with coverage frequencies. In each task, we compare the performance of our method with the other tests of independence mentioned in the introduction. 

We generate the simulated samples as follows. Let $\Sigma=\begin{pmatrix}1&0.5&0.5\\0.5&1&0.5\\0.5&0.5&1\end{pmatrix}$. Independently draw $\bm{P}_{raw}$ and $\bm{W}_{raw}$ from $\mathcal{N}(0,\Sigma)$. Then for $k=1,2,3$, construct $P_{k}$ by $\Phi(P_{raw,k})+1$ and $W_{k}$ by $\Phi(W_{raw,k})$, where $\Phi$ is the cumulative distribution function of the standard normal distribution. This ensures $P_{k}>W_{k}>0$ with probability 1 and thus the optimal demand is an interior solution. We then generate $Y_{k}$ by the first order condition $Y_{k}=\theta_{k}/(P_{k}-W_{k})$, where $(\theta_{1},\theta_{2},\theta_{3})=(0.2,0.3,0.5)$.

\subsubsection{Rejection Frequencies}
We first compare the rejection frequencies for Chatterjee's $\xi$, Hoeffding's $D$ \citep{hoeffding1948non}, Blum-Kiefer-Rosenblatt's $R$ \citep{blum1961distribution} and Bergsma-Dassios-Yanagimoto's $\tau^{*}$ \citep{bergsma2014consistent,yanagimoto1970measures}. We set $\alpha=0.1$. We consider two null hypotheses, one equal to the true value $(0.2,0.3,0.5)$, whereas the other equal to $(0.2+0.1/3, 0.3+0.1/3, 0.5-0.2/3)$.

For Chatterjee's $\xi$, we reject if there is at least one $k$ such that $\sqrt{n}\xi_{n}(P_{k}-\theta_{k}^{0}/Y_{k},P_{k})/\sqrt{0.4}> z_{(1-\alpha)^{1/3}}$ where $\bm{\theta}^{0}\coloneqq(\theta_{k}^{0})$ is the null hypothesis and the critical value is implied by Theorem \ref{thm.joint1}. The value of $\xi_{n}$ under each $\bm{\theta}^{0}$ is computed using the R package \texttt{XICOR} \citep{chatterjee2020xicor}.

For the other tests, since their joint limiting distributions are unavailable to the best of our knowledge, we do Bonferroni correction by using the critical values at $\alpha/3$ for each $\theta_{k}$. Computationwise, we use the R package \texttt{independence} \citep{even2020independence}. The algorithms the package adopts achieve $O(n\log n)$ computation time for all these tests.

Table \ref{tab.power.prod3} reports the results. The rejection frequencies are computed based on 500 simulation replications. When the null is equal to the true parameter value, the rejection frequencies of Chatterjee's $\xi$ is close to the nominal level $0.1$ for all three sample size. The other methods are less stable. When the null is not equal to the true value, the rejection frequency of Chatterjee's $\xi$ is lower than the others when $n=200$; this is expected as Chatterjee's $\xi$ has lower power than the other three as documented in \cite{shi2022power}. However, this difference becomes much smaller at $n=1,000$. At $n=5,000$, all four tests reject the false hypothesis in all 500 replications. 

\begin{table}[htbp]
  \centering
  \caption{Rejection Frequencies; $\alpha=0.1$}
    \begin{tabular}{cccccc}
    \toprule
    \toprule
       $n$  &Null Hypothesis &  $\xi$ & $D$&  $R$ & $\tau^{*}$ \\
          \midrule
    \multirow{2}{*}{200}& $(0.2,0.3,0.5)$ &0.092 &0.122   &0.112& 0.114 \\  
                        & $(0.7/3,1/3,1.3/3)$       & 0.424 &0.696 &0.676& 0.682  \\     
                        \midrule
    \multirow{2}{*}{1,000}& $(0.2,0.3,0.5)$&  0.102& 0.068  &0.066&0.068 \\  
                        & $(0.7/3,1/3,1.3/3)$  & 0.960&  1.000 & 1.000&1.000\\ 
                      \midrule  
    \multirow{2}{*}{5,000}& $(0.2,0.3,0.5)$ &0.098 &0.110   &0.108& 0.110 \\  
                        & $(0.7/3,1/3,1.3/3)$        & 1.000&1.000 &1.000& 1.000  \\     
                        \bottomrule
    \end{tabular}%
  \label{tab.power.prod3}%
\end{table}%

\subsubsection{Confidence Intervals}
For Chatterjee's $\xi$, we construct a $(1-\alpha)$ joint confidence set for $\bm{\theta}$ by equation \eqref{eq.CSonly}, treating $P_{k}$ as $Z_{k}$; specifically, for each $k$, we search over a grid of 1,000 nodes in $(0,1)$, denoted by $\Theta$, to find points that are in the set defined in \eqref{eq.CSindi}, and then take the minimum and the maximum. For the other tests, we invert the individual tests with the $\alpha/3$ critical value by a similar grid search.

Table \ref{tab.CI} shows the results averaged over 500 repetitions. The lower and upper bounds (LB and UB) of our confidence intervals shrink toward the true parameters as the sample size increases. Compared to the confidence intervals obtained by inverting the other tests, our intervals have longer lengths, but the differences in length are small especially when the sample size is large. 

In terms of computation time, on the other hand, our method is about 1,000 times faster than the other three. This suggests that when the sample size (and the number of goods $K$, since computation time increases linearly in $K$ for all these methods) is large, our method will be much more efficient in computation while maintaining reasonably large power. 

\begin{table}[H]
\centering
\caption{Confidence Intervals and Computation Time with $\alpha=0.1$}\label{tab.CI}
\begin{tabular}{ccccccc}
\toprule
\toprule
&&& $\xi$&$D$& $R$&$\tau^{*}$\\
\midrule
\multirow{10}{*}{$n=200$}&\multirow{3}{*}{$\theta_{1}=0.2$} &LB& 0.146&0.171&0.171&0.171\\
&&UB&0.360 &0.242&0.242&0.242 \\
&&Length&0.214 &0.070&0.071&0.071 \\
\cmidrule{2-7}
&\multirow{3}{*}{$\theta_{2}=0.3$} &LB&0.217 &0.257&0.256&0.256\\
&&UB& 0.551 &0.363&0.364&0.364 \\
&&Length& 0.334&0.107&0.108&0.107 \\
\cmidrule{2-7}
&\multirow{3}{*}{$\theta_{3}=0.5$} &LB&0.359 &0.428&0.427&0.427\\
&&UB&0.850 &0.605&0.606& 0.606\\
&&Length&0.492 &0.177&0.179& 0.178\\
\cmidrule{2-7}
&\multicolumn{2}{c}{Time (in seconds)}&0.28&271.10&165.21&207.90\\
\midrule
\multirow{10}{*}{$n=1000$}&\multirow{3}{*}{$\theta_{1}=0.2$} &LB&0.172 &0.187&0.187&0.187\\
&&UB&0.246 &0.217&0.217&0.217 \\
&&Length&0.074 &0.029&0.030& 0.030\\
\cmidrule{2-7}
&\multirow{3}{*}{$\theta_{2}=0.3$} &LB&0.255&0.280&0.280&0.280\\
&&UB&0.371 &0.324&0.325& 0.324\\
&&Length& 0.116&0.045&0.045& 0.045\\
\cmidrule{2-7}
&\multirow{3}{*}{$\theta_{3}=0.5$} &LB& 0.424&0.465&0.465&0.465\\
&&UB&0.623 &0.540&0.540&0.540 \\
&&Length&0.198 &0.075&0.075& 0.075\\
\cmidrule{2-7}
&\multicolumn{2}{c}{Time (in seconds)}&0.80&1144.78&792.28&868.39\\
\midrule
\multirow{10}{*}{$n=5000$}&\multirow{3}{*}{$\theta_{1}=0.2$} &LB&0.187 &0.194&0.194&0.194\\
&&UB&0.216&0.207&0.207& 0.207\\
&&Length& 0.028&0.012&0.012&0.012 \\
\cmidrule{2-7}
&\multirow{3}{*}{$\theta_{2}=0.3$} &LB& 0.280&0.291&0.291&0.291\\
&&UB&0.325&0.310 &0.310&0.310 \\
&&Length& 0.046&0.019&0.019& 0.019\\
\cmidrule{2-7}
&\multirow{3}{*}{$\theta_{3}=0.5$} &LB&0.464 &0.485&0.485&0.485\\
&&UB& 0.546&0.517&0.517&0.517 \\
&&Length& 0.082&0.032&0.032& 0.032\\
\cmidrule{2-7}
&\multicolumn{2}{c}{Time (in seconds)}&3.89&3928.21&3566.47&3702.90\\
\bottomrule
\end{tabular}
\end{table}

\subsubsection{Welfare Loss Bounds}
For the welfare loss, we consider a hypothetical price change $\bm{\Delta}=(0.5,0.8,0.2)$ with consumption fixed at $(0.2,0.6,0.8)$.  
We maximize the standardized welfare loss function \eqref{eq.objst} under one of the four constraints: 
  \begin{itemize}
    \item[i)] Confidence intervals and $\sum_{k}\theta_{k}=1$;
    \item[ii)] Confidence intervals only;
    \item[iii)] $\theta_{k}\in [10^{-6},1]$ and $\sum_{k}\theta_{k}=1$;
    \item[iv)] $\theta_{k}\in [10^{-6},1]$ only.
  \end{itemize}
For constraints ii) and iv), we maximize equation \eqref{eq.objst} simply by evaluating it at the upper bounds of the $\theta_{k}$s. For constraints i) and iii), we use the \texttt{CVXR} package for R \citep{CVXR} to conduct optimization. See Appendices \ref{appx.implement} for details. Note that the maximum loss under iii) and iv) are constant across different sample sizes. The true value of the standardized welfare loss is 0.525.

\begin{table}[h]
\centering
\caption{Upper Bounds on Standardized Welfare Loss with $\alpha=0.1$ and Coverage Frequencies; True Value $=0.525$}\label{tab.wel}
\begin{tabular}{ccccccccccc}
\toprule
\toprule
& \multicolumn{2}{c}{$\xi$}&\multicolumn{2}{c}{$D$}&\multicolumn{2}{c}{$R$}&\multicolumn{2}{c}{$\tau^{*}$}&\multicolumn{2}{c}{Data Free}\\
\midrule
 &i)&ii)&i)&ii)&i)&ii)&i)&ii) &iii) & iv)\\
 \midrule
 \multicolumn{11}{c}{\textit{Panel A. Upper Bounds}}\\
 $n=200$     & 0.561 &0.598 &0.540&0.550&0.540&0.551&0.540&0.550&\multirow{3}{*}{0.574} &\multirow{3}{*}{0.659}\\
 $n=1000$     & 0.544 &0.554&0.532&0.536&0.532&0.536&0.532&0.536&&\\
 $n=5000$   & 0.533 &0.536 & 0.528&0.530&0.528&0.530&0.528&0.530&&\\
  \midrule
 \multicolumn{11}{c}{\textit{Panel B. Coverage Frequencies}}\\
 $n=200$     & 1.000 &1.000 &0.968&0.986&0.970&0.986&0.970&0.986&\multirow{3}{*}{1.000} &\multirow{3}{*}{1.000}\\
 $n=1000$     & 1.000 &1.000&0.992&0.998&0.992&0.998&0.992&0.998&&\\
 $n=5000$   & 1.000 &1.000 & 0.978&0.992&0.978&0.992&0.978&0.992&&\\
 \bottomrule
 \end{tabular}
 \end{table} 

Table \ref{tab.wel} shows the upper bounds on the standardized welfare loss, as well as the coverage frequencies of these bounds, averaged over 500 simulation repetitions\footnote{For $D$, $R$, and $\tau^{*}$, there are up to three simulation repetitions in which constraints i) yields an empty set because there are no values in the confidence intervals satisfying $\sum_{k}\theta_{k}=1$. The upper bounds and the coverage frequencies for these three methods are thus averaged over the rest of the repetitions.}. We have the following four findings. i) All the upper bounds shrink towards the true welfare loss as the sample size increases. ii) Information from data does improve the bounds compared to bounds without using data, as the bounds under constraint i) (or ii)) are smaller than those under constraint iii) (or iv)). iii) The standardized welfare loss bounds under $\xi$ are larger than those under the other three statistics, but the differences are small and shrink toward zero as the sample size increases. iv) The coverage frequencies of these upper bounds are always close or equal to one, greater than $(1-\alpha)$. This is expected because Theorem \ref{thm.welfare} shows that the coverage probability for the standardized welfare loss bounds by construction can be greater than the coverage probability for the parameters.

\subsection{Scalability When There Are Many Goods}
In this section, we examine the performance and computation time of our confidence set and welfare loss bounds when there are many goods.
We set the number of goods $K$ equal to $10,50$ or $100$. Parameters $\theta_{k}$s are $K$ equally spaced numbers between $0.1$ and $0.9$. 
For each good $k$, $P_{k}$ and $W_{k}$ are independently drawn from $\text{Unif}[1,2]$ and $\text{Unif}[0,1]$, respectively. Consumption $Y_{k}$ is constructed by the first-order condition. 

We start by examining the rejection frequency of the joint test. We consider two null hypotheses. The first hypothesis $\bm{\theta}^{0}_{1}$ is equal to the true value. For the second hypothesis $\bm{\theta}^{0}_{2}$, $\theta_{2,k}^{0}=\theta_{k}+0.03$ if $k\leq K/2$ and $\theta_{2,k}^{0}=\theta_{k}-0.03$ if $k> K/2$. Computation is done similarly as Section \ref{sec.MC.3} with critical value equal to $z_{(1-\alpha)^{1/K}}$, where $\alpha=0.1$.

\begin{table}[htbp]
  \centering
  \caption{Rejection Frequencies; $\alpha=0.1$}
    \begin{tabular}{ccccccc}
    \toprule
    \toprule
          & \multicolumn{2}{c}{$K=10$} & \multicolumn{2}{c}{$K=50$} & \multicolumn{2}{c}{$K=100$} \\
       \midrule
         $n$   &$\bm{\theta}^{0}_{1}$& $\bm{\theta}^{0}_{2}$  &$\bm{\theta}^{0}_{1}$& $\bm{\theta}^{0}_{2}$  &$\bm{\theta}^{0}_{1}$& $\bm{\theta}^{0}_{2}$ \\
          \midrule
    200 & 0.114 &0.426 & 0.112 & 0.482 &0.088& 0.536\\
    1000 & 0.096 &0.966 & 0.078 & 0.998 &0.108& 1.000 \\
    5000 & 0.084 &1.000& 0.100 & 1.000 &0.102& 1.000  \\
    \bottomrule
    \end{tabular}%
  \label{tab.manygoods.rej}%
\end{table}%
Table \ref{tab.manygoods.rej} shows the rejection frequencies in 500 simulation repetitions. At the true value, the rejection frequencies are close to the nominal level $0.1$. Under the false hypothesis, the rejection frequencies increase with the sample size and are close to or equal to 1 when $n$ reaches 1,000.

Next, we compute the $(1-\alpha)$ confidence sets by \eqref{eq.CSonly}, treating $Z_{k}=P_{k}$ and using grid search as before. For the bounds on the standardized welfare loss, we only consider constraint ii) in Section \ref{sec.MC.3}. We draw the hypothetical price change $\bm{\Delta}$ and consumption level $\bm{y}^{0}$ from $\text{Unif}[0.1,1.1]$ and $\text{Unif}[0.2,1.2]$ once, respectively; for each $K$, they stay the same across all simulation replications and all sample sizes. We compute the upper and lower bounds on the standardized welfare loss by substituting the upper and lower bounds of the confidence intervals of the $\theta_{k}$s, respectively.

Table \ref{tab.manygoods} presents the results averaged over 500 simulation replications. Numbers in columns ``Length'' are the average length of the $K$ confidence intervals. The confidence intervals shrink as the sample size increases for all $K$. Columns ``Welfare'' demonstrate the upper and lower bounds on the standardized welfare loss, the true values of which are shown as the numbers in parentheses. The bounds are tight even when the number of goods reaches 100 while the sample size is only 200. Columns ``Coverage'' are frequencies that these bounds cover the true standardized welfare loss, which are always 1 across all setups.
\begin{sidewaystable}[htbp]\setlength{\tabcolsep}{2pt}
  \centering
  \caption{Confidence Sets, Computation Time and Standardized Welfare Loss with Many Goods; $\alpha=0.1$}
    \begin{tabular}{cccccccccccccccc}
    \toprule
    \toprule
          & \multicolumn{5}{c}{$K=10$} & \multicolumn{5}{c}{$K=50$} & \multicolumn{5}{c}{$K=100$} \\
       \midrule
          & \multicolumn{2}{c}{CI} & \multicolumn{3}{c}{Welfare (1.34)} & \multicolumn{2}{c}{CI} & \multicolumn{3}{c}{Welfare (2.81)} & \multicolumn{2}{c}{CI} & \multicolumn{3}{c}{Welfare (4.15)} \\
          \midrule
       $n$   & Length & Time (s) & Lower& Upper&Coverage& Length & Time (s) & Lower& Upper&Coverage & Length & Time (s) & Lower& Upper&Coverage \\
          \midrule
    200 & 0.43 &1.01 & 1.19 & 1.54 & 1& 0.50 & 5.05 & 2.48 & 3.21 &1& 0.52 & 10.35 &  3.62 &4.86&1\\
    1000 & 0.21 &2.77 & 1.27 & 1.43 &1& 0.25 & 13.81& 2.64 & 3.01 &1& 0.26 & 27.62& 3.88 & 4.48 &1\\
    5000 & 0.10 & 13.51& 1.31 & 1.38 &1& 0.11 & 59.17& 2.73 & 2.89 &1& 0.12 & 134.31& 4.03 &4.29&1 \\
    \bottomrule
    \end{tabular}%
  \label{tab.manygoods}%
\end{sidewaystable}%

Table \ref{tab.manygoods} also demonstrates the computation time (in seconds) for confidence intervals\footnote{The average computation time for the standardized welfare loss bound is always below $10^{-4}$ seconds and thus suppressed. Note that the objective function \eqref{eq.objst} does not depend on data, so sample size does not affect computation.} under various combinations of $n$ and $K$. By simple calculation, one can see that the computation time is approximately linear in $K$ for each $n$. This is expected because we calculate the confidence set for each good independently. As an implication, parallel computation is feasible to keep the computation time constant across $K$, and thus the problem is scalable. 

\subsection{The Intersection Approach and Rejection Probability}\label{sec.mc.data}
In this simulation exercise, we examine the performance of the intersection approach. Different from the previous sections, the data-generating process is now designed to match some key moments in the food consumption data set in Section \ref{sec.food}. 
Specifically, we utilize the sample extract from the Stanford Basket Dataset, a household-level scanner panel data set,  created by \cite{pump} for the empirical results presented in Table 4 of their paper. There are two goods in the sample: Ice cream and other foods. 

We first construct a data set with valid observations. For each of the 494 households in the sample, there are 26 observations: One observation corresponds to a period of 4 weeks over a two-year span.
In Section \ref{sec.food}, we will perform household-level welfare analysis using the intersection approach for households with at least 19 periods of strictly positive consumption of ice cream and other foods. When $\alpha=0.1$, the intersection approach yields nonempty confidence intervals for 38 households. In this section, for simplicity, we pool all these 38 households together and obtain a data set of 763 observations.   

We then design the following data-generating process to match the sample mean and variance of the observed prices and consumption in the pooled data set. For each of the two consumption goods, we assume its price follows a truncated normal distribution. The lower and upper bounds are set to equal the observed minimum and maximum of the price data. For the pre-truncation mean and variance, we calibrate them by grid search so that the resulting post-truncation mean and variance match those of the observed price data.

For the unobservables and consumption, we first draw $(W_{raw,1},W_{raw,2})$ from $\mathcal{N}(0,\Sigma_{W})$ where $\Sigma_{W}=\begin{pmatrix}
    1&\rho_{W}\\\rho_{W}&1
\end{pmatrix}$ and $\rho_{W}\in\{0.1,0.7\}$. Then to guarantee $0<W_{k}<P_{k}$ with probability 1, we construct $W_{1}=0.553\Phi(W_{raw,1})^{2}$ and $W_{2}=1.1596\Phi(W_{raw,2})^{2}$. Finally, we construct $Y_{k}=\theta_{k}/(P_{k}-W_{k})$ with $(\theta_{1},\theta_{2})=(1.1,0.6)$. 

Table \ref{tab.moments} presents the sample mean and variance of the observed and simulated prices and consumption averaged across 1,000 simulation replications; in each simulation replication, the sample size is 5,000. We can see that our data-generating process matches these moments well.

    \begin{table}[htbp]
  \centering
  \caption{Moments of the Data and Simulated Samples}
    \begin{tabular}{cccccc}
    \toprule
    \toprule
        & & $P_{1}$& $P_{2}$&$Y_{1}$&$Y_{2}$  \\
                 \midrule
     \multirow{2}{*}{Mean}& Data &0.745 &1.439 & 2.323& 0.709 \\
      &    Simulation&0.745 &1.439 & 2.311& 0.723\\
               \midrule
     \multirow{2}{*}{Variance}& Data &0.004 &0.007 &1.593 &0.284 \\
     &      Simulation& 0.004& 0.007& 1.701 & 0.258\\
                        \bottomrule

    \end{tabular}%
  \label{tab.moments}%
\end{table}%

We now turn to our main simulation results. We first examine the joint rejection frequencies of three different approaches: Chatterjee's $\xi$, SUR and the delta method, and the intersection approach. For Chatterjee' $\xi$, since there are two goods, we calculate the rejection frequencies similar to Sections \ref{sec.MC.3} with critical value $ z_{\sqrt{1-\alpha}}$. For SUR, we calculate the rejection frequencies based on the sup-t test. The asymptotic rejection probability is also exactly $\alpha$ under the null. For the intersection approach, we reject if and only if at least one of the tests based on Chatterjee and SUR suggests rejection; the size of each test is adjusted to $1-\sqrt{1-\alpha}$ so that the joint rejection probability under the null is controlled at $\alpha$ based on Theorem \ref{thm.joint2}. 

\begin{table}[htbp]
  \centering
  \caption{Rejection Frequences; True Parameter$=(1.1,0.6)$; $\alpha=0.1$}
    \begin{tabular}{cccccc}
    \toprule
    \toprule
   &&    \multicolumn{2}{c}{Low Correlation ($\rho_{W}=0.1$)}& \multicolumn{2}{c}{High Correlation ($\rho_{W}=0.7$)} \\
\cmidrule{3-6}
       $n$  &Method &  $(1.1,0.6)$ & $(1.3,0.8)$&  $(1.1,0.6)$ & $(1.3,0.8)$ \\
          \midrule
    \multirow{3}{*}{500}& $\xi$ only  &0.106 &0.912   &0.082& 0.934 \\  
                        & SUR only        & 0.106 &0.714 &0.112& 0.842  \\ 
                        & Intersection &  0.134 & 0.936& 0.130& 0.969   \\           
                        \midrule
    \multirow{3}{*}{1,000}& $\xi$ only&  0.108& 0.998  &0.102&0.998 \\  
                        & SUR only      & 0.088&  0.854 & 0.096&0.958\\ 
                        & Intersection &  0.096&  0.996 & 0.108&  0.996 \\   
                        \bottomrule

    \end{tabular}%
  \label{tab.power}%
\end{table}%
Table \ref{tab.power} presents the results based on 500 simulation replications. We consider two hypotheses, the true value $(1.1,0.6)$ and the false value $(1.3,0.8)$. The results show that all three methods, under the true parameter value, have rejection frequencies close to $\alpha=0.1$. This verifies the asymptotic independence results in our Theorems \ref{thm.joint1} and \ref{thm.joint2}.  Under the false hypothesis, these methods all have large power, and the power increases as the sample size increases. In particular, SUR and the intersection approach improve when the unobservables are more correlated. 

Finally, we compute the standardized welfare loss bounds based on the confidence intervals obtained by the three methods. Again, we treat $Z_{k}=P_{k}$ for all $k$. When using Chatterjee's $\xi$ alone, the confidence sets are constructed following \eqref{eq.CSindi} and \eqref{eq.CSonly} by grid search in $\Theta_{k}=[10^{-6},2]$ with 2,000 grid points. For the intersection approach, grid search with the same number of grid nodes is conducted in the sup-t confidence interval following \eqref{eq.CSindiinter} and \eqref{eq.CSinter}. The consumption level is fixed at the sample median of the observed consumption. The price increase is 20\% of the sample median of the observed prices.

Table \ref{tab.welfare.intersect} presents the standardized welfare loss bounds and the length between the lower and upper bounds averaged over 500 simulation replications. We can see that the intersection approach always achieves the shortest length. Due to the efficiency gain from SUR, the improvement compared to using $\xi$ alone is larger when the unobservables are more correlated. 

The table also shows the frequencies of the true standardized welfare loss covered by the confidence bounds. It is always close or equal to one, greater than $(1-\alpha)$, similar to the previous sections.
\begin{table}[htbp]
  \centering
  \caption{Standardized Welfare Loss Bounds; True Value $=1.266$}
    \begin{tabular}{cccccccccc}
    \toprule
    \toprule
     &   &   \multicolumn{4}{c}{Low Correlation ($\rho_{W}=0.1$)} & \multicolumn{4}{c}{High Correlation ($\rho_{w}=0.7$)}  \\
       \midrule
       $n$ & Method & Lower & Upper & Length &Coverage &   Lower & Upper & Length &Coverage \\
         \midrule
    \multirow{3}{*}{500} & $\xi$ only   & 1.231& 1.320& 0.090& 1.000 & 1.230 &1.321 &0.091& 1.000 \\  
                          & SUR only        & 1.210  &1.297  &0.087&0.990& 1.229& 1.291 &0.062   &0.988 \\ 
                          & Intersect  &  1.225 &1.297&0.072&0.992 & 1.231&  1.291 &0.060 &0.994\\          
                        \midrule

       \multirow{3}{*}{1000}  & $\xi$ only   & 1.248& 1.297& 0.050&1.000 & 1.248 &1.298 &0.050 &1.000 \\  
                          & SUR only         & 1.232  &1.291  &0.059&0.988& 1.242& 1.285 &0.043 & 0.992  \\ 
                          & Intersect  &  1.243 &1.289&0.046 & 0.992&1.246&  1.285 &0.039 &0.994\\                     
                        \bottomrule

    \end{tabular}%
  \label{tab.welfare.intersect}%
\end{table}%

\section{Empirical Applications}\label{sec.application}
\subsection{Demand for Gasoline}\label{sec.gas}
We study the standardized welfare loss of a hypothetical gasoline price increase. We take the data set constructed by \cite{blundell2017nonparametric} using the household-level 2001 National Household Travel Survey (NHTS). The sample contains 3,640 observations with annual gasoline demand ($\tilde{Y}$), price of gasoline ($P$) and the distance between one of the major oil platforms in the Gulf of Mexico and the state capital\footnote{In the data set, the distance takes on 34 values ranging from 0.361 to 3.391. We treat it as a continuous variable. In Appendix \ref{appx.gas}, we further smooth it by adding a small noise term to it. The results are almost the same.} ($Z$; see \cite{blundell2017nonparametric} for a more detailed discussion). We consider the daily demand $Y$ by dividing $\tilde{Y}$ by 365.

Under our utility function \eqref{eq.utility} and the first order condition \eqref{eq.foc} with $K=1$, we can construct confidence intervals for $\theta$ using our method, 2SLS and the delta method, and the intersection approach. When using Chatterjee's $\xi$ alone, our confidence interval follows \eqref{eq.CSindi} by grid search with 5,000 grid nodes over $\Theta=[10^{-6},6]$. For the intersection approach \eqref{eq.CSindiinter}, grid search is done over the $\sqrt{1-\alpha}$ confidence interval obtained by 2SLS and the delta method. Table \ref{tab.ci.gas} presents the results for $\alpha=0.05$. For the standardized welfare loss bounds, we let the price of gasoline increase by twice of the standard deviation (0.076). The current demand is set at the $u$-th sample quantile ($u\in (0,1)$) of $Y$, $q_{Y}(u)$, for $u=0.9,0.5$ and $0.1$. 

\begin{table}[H]
  \centering
  \caption{Gasoline Demand: Confidence Interval and Standardized Welfare Loss Bounds}
    \begin{tabular}{ccccc}
    \toprule
    \toprule
    &Confidence Intervals&\multicolumn{3}{c}{Standardized Welfare Loss Bounds}\\ 
    \midrule
          & &{$q_{Y}(0.9)$}& {$q_{Y}(0.5)$}&{$q_{Y}(0.1)$} \\
    \midrule
    2SLS only & $[0.842, 4.142]$ & $[4.584,6.325]$ & $[2.771,3.374]$ & $[1.404,1.552]$\\
    $\xi$ only & $[1.438, 6]$ & $[5.312,6.545]$ & $[3.046,3.438]$ & $[1.476,1.565]$ \\
      Intersection  & $[1.438 , 4.374]$ &$\bm{[5.312,6.362]}$ & $\bm{[3.046,3.385]}$ & $\bm{[1.476,1.554]}$\\
    \bottomrule
    \end{tabular}%
  \label{tab.ci.gas}%
\end{table}%
From Table \ref{tab.ci.gas}, the intersection approach improves the lower bound of the confidence interval a lot compared to the lower bound under 2SLS alone, whereas the upper bound only increases a bit. See Appendix \ref{appx.shape} for more discussion. Similarly, the standardized welfare loss bounds are tightest under the intersection approach (entries in boldface). In particular, the difference between the upper and lower bounds shrinks by about 40\%, 44\%, and 47\% when intersecting the two confidence sets compared to using 2SLS alone for the three demand levels, respectively. {Table \ref{tab.ci.gas} shows that a price increase leads to more severe damage to the welfare of individuals who consume more gas, which is reasonable.}

\subsubsection{A Varying Coefficient Model}

One limitation of the quasilinear utility function is the missing income effect. To address this issue, we now allow $\theta$ to depend on income, $I$. Let $med(I)$ be the median of the income distribution. Random variable $\theta^{I}=\theta^{H}$ if $I>med(I)$ and  $\theta^{I}=\theta^{L}$ if $I\leq med(I)$. Assume that the instrument is independent of the unobservable conditional on income. In this section, we compute the welfare loss bounds in the two subsamples defined by whether the individual's income is greater than the sample median. Similar to the case of a homogeneous $\theta$, the intersection approach largely improves the lower bound of the welfare loss in both subsamples compared to using 2SLS alone. To save space, we only report the results obtained by this approach. Again, $\alpha=0.05$. 

\begin{table}[H]
  \centering
  \caption{Confidence Interval and Standardized Welfare Loss with an Income Dependent Parameter}
    \begin{tabular}{ccccc}
    \toprule
    \toprule
    &Confidence Intervals&\multicolumn{3}{c}{Standardized Welfare Loss Bounds}\\ 
    \midrule
          & &$q_{Y}(0.9)$& $q_{Y}(0.5)$&$q_{Y}(0.1)$\\
    \midrule
    $\theta^{H}$ & $[1.463, 7.033]$ & $[5.641,  7.128]$ & $[3.388, 3.904]$ & $[1.779, 1.917]$\\
    $\theta^{L}$ & $[1.026,  2.876]$ & $[4.598, 5.624]$ & $[2.599,  2.922]$ & $[1.207, 1.276]$ \\
    \bottomrule
    \end{tabular}%
  \label{tab.ci.gas.random}%
\end{table}%

Table \ref{tab.ci.gas.random} suggests heterogeneity in $\theta$ and consequently in the standardized welfare loss. From the results, individuals with higher incomes tend to have a larger $\theta$. This is reasonable because higher-income individuals may rely more on driving than public transportation, so the marginal utility of consuming one more unit of gas, captured by $\theta$, is higher. Since they value gas more than those with lower income, the welfare of higher income individuals suffers more when facing an increase in gas price, coherent with the estimated standardized welfare loss bounds.

\subsection{Food Demand}\label{sec.food}

In this application, we consider food demand using the Stanford Basket Dataset. As mentioned in Section \ref{sec.mc.data}, it is a household-level scanner panel data set, and we use the sample extract created by \citet{pump}. Two goods, ice cream and other foods, are in the sample. 

 We conduct household-level welfare analysis taking advantage of the panel data structure. Recall that there are 26 observations for each of the 494 households in the sample.
We focus on 41 households that have at minimum 19 periods of strictly positive consumption of ice cream and other foods; each household can have a different $\bm{\theta}$.   
We use the last observation for the counterfactual welfare analysis and the remaining observations for the construction of confidence sets. The i.i.d. assumption required in our theory may be strong here; we regard it as a convenient approximation.

\begin{figure}[htbp]
\begin{center}
\includegraphics[scale=0.5]{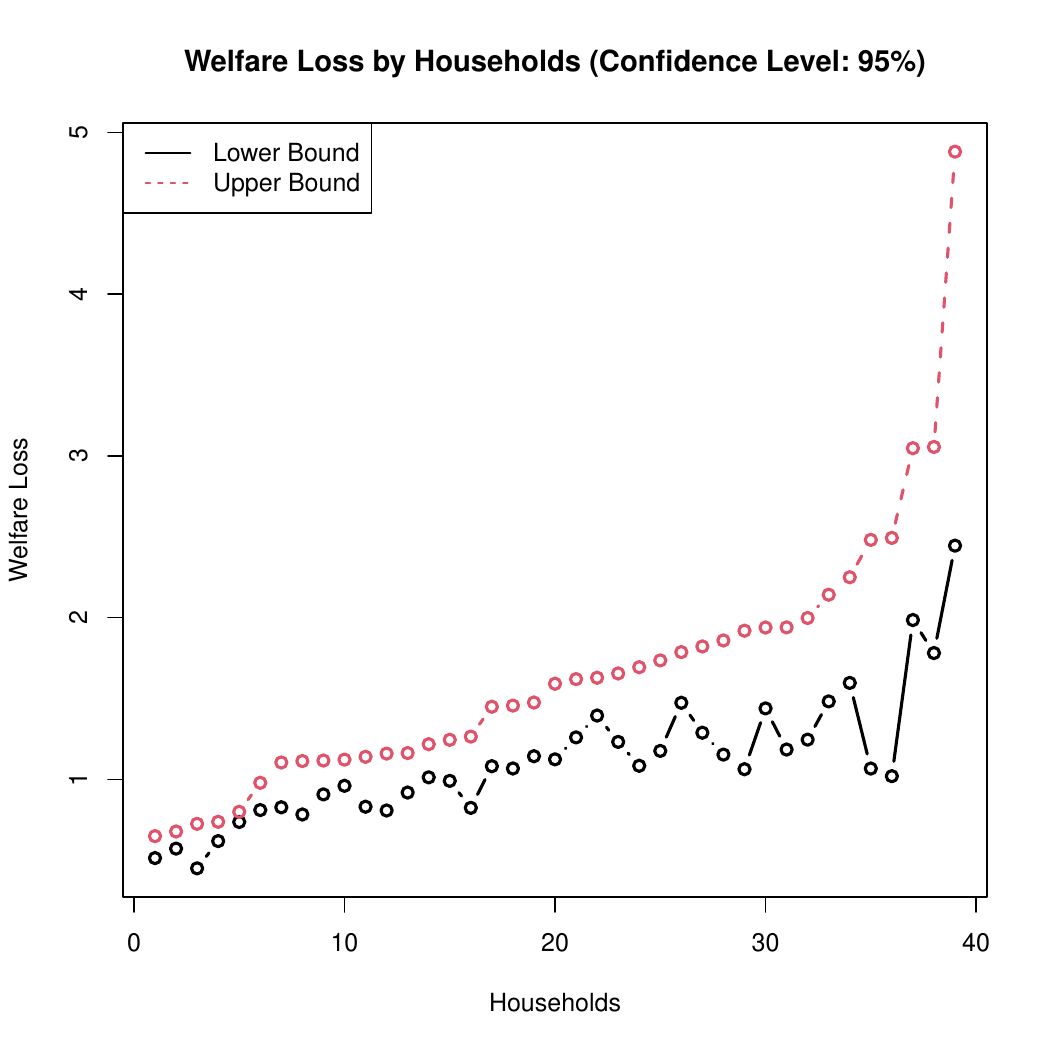}
\caption{Household-Level Standardized Welfare Loss Bounds (Confidence Level $=0.95$)}
\label{fig.pump}
\end{center}
\end{figure}

We assume the prices are exogenous. To construct a $(1-\alpha)$ confidence set where $\alpha=0.05$, we first construct a $\sqrt{1-\alpha}$ sup-t confidence band for each household by SUR and the delta method. We then do a grid search with 5,000 grid nodes within the resulting confidence intervals as equations \eqref{eq.CSindiinter} and \eqref{eq.CSinter}. Two out of 41 households have an empty confidence set. 

For the welfare analysis for the 39 households with nonempty confidence sets, we compute the standardized welfare loss bounds in the same way as before. Figure~\ref{fig.pump} shows the lower and upper bounds on the standardized welfare loss for the 39 households when both prices increase by 10\%. 
The maximal standardized welfare loss of the households is sorted in ascending order.
Household heterogeneity is visible and the lower and upper bounds are relatively tight for most households.\footnote{In Appendix \ref{appx.food}, we test one assumption in our model \eqref{eq.demand} that the demand for a good is only affected by its own price. This hypothesis is rejected for only 3 of the 39 households. The figure excluding these 3 households is presented in that appendix; it has a very similar shape as Figure \ref{fig.pump}. We thank an anonymous referee for suggesting this test.}

\section{Conclusion}
In this paper, we propose a novel framework for individual-level welfare analysis. At any desired confidence level, our method can compute the bounds on the standardized welfare loss under a price increase for every individual in a sample. The inferential method is computationally simple and scalable by solving a simple scalable optimization problem constrained by the confidence set for the parameters in the model. 

We also propose a new method to construct confidence sets under independence based on the new asymptotic results of Chatterjee's test of independence developed in this paper; the new method is easy to compute and robust to nonlinearity, weak instruments, and partial identification; these results may be of independent interest. To sharpen the confidence set, we propose to intersect our confidence set with an alternative one when available. As in our second empirical example, the intersection approach sometimes yields an empty set. In that case, we may try a very small $\alpha$, for instance, 0.5\%. If the intersection becomes nonempty at the new $\alpha$, we can report the new confidence interval as auxiliary information. In our empirical example, all the households have nonempty confidence sets under the interaction approach when we set $\alpha=1\%$. If, on the other hand, the intersection is still empty under the new $\alpha$, then it is a warning that the model may be misspecified. Misspecification is beyond the scope of this paper, and we leave it to future work.

Another direction of future research would be to apply our framework to demand models such as \cite{dubois2014prices} and \cite{allcott2019food} to conduct individual-level welfare analysis. 

\appendix
\section{On the Shape and Size of the Confidence Set}\label{appx.shape}

In this appendix, we provide heuristics about how $\xi_{n}(P,Z)$ and $\xi_{n}(Y,Z)$ affect the shape and size of our confidence set. We first introduce the following lemma.
\begin{lem}\label{lem.monoinv}
Let $f$ be a strictly monotonic function. For two random variables $R^{(1)}$ and $R^{(2)}$, suppose we have an i.i.d. sample of them where there are no ties, then $\xi_{n}(f(R^{(1)}),R^{(2)})=\xi_{n}(R^{(1)},R^{(2)})$.
\end{lem}
\begin{proof}
Let $r_{i}^{0}$ and $r_{i}^{f}$ be the rankings of $R^{(2)}_{i}$ after the data are sorted by $R^{(1)}$ and $f(R^{(1)})$ in ascending order, respectively. If $f$ is strictly increasing, $r_{i}^{0}=r_{i}^{f}$ for all $i$; we are done. If $f$ is strictly decreasing, the order is reversed so $r_{i}^{f}=r_{n+1-i}^{0}$. Then by letting $n-i=j$,
\begin{align*}
\sum_{i=1}^{n-1}|r^{f}_{i+1}-r^{f}_{i}|=&\sum_{i=1}^{n-1}|r^{0}_{n-i}-r^{0}_{n+1-i}|\\
=&\sum_{j=n-1}^{1}|r^{0}_{j}-r^{0}_{j+1}|\\
=&\sum_{i=1}^{n-1}|r^{0}_{i+1}-r^{0}_{i}|.
\end{align*}
The desired result is obtained in view of equation \eqref{eq.xi}.
\end{proof}
 \begin{appxrem}
 The conclusion in Lemma \ref{lem.monoinv} still holds when the $R^{(2)}_{i}$s have ties. It follows the same proof since the denominator in the formula of $\xi_{n}$ in footnote 1 is invariant to reordering.
 \end{appxrem}
Using our demand model with one good as an example, our inferential method focuses on the statistic $\xi_{n}(P-\theta/Y,Z)$ where $Z$ is a continuous instrument. Denote the true parameter by $\theta_{0}>0$. Our model says $P-\theta_{0}/Y=W$ so for an arbitrary $\theta$, $P-\theta/Y=(1-\theta/\theta_{0})P+(\theta/\theta_{0})W$. Let $c_{1-\alpha}=\sqrt{0.4}z_{1-\alpha}$. Then our $(1-\alpha)$ confidence set \eqref{eq.CSindi} is equal to
\[
 \left\{\theta\geq 0: \sqrt{n}\xi_{n}\left(\left(1-\frac{\theta}{\theta_{0}}\right)P+\frac{\theta}{\theta_{0}}W,Z\right)\leq c_{1-\alpha}\right\}.
\] 

First, when $\theta=\theta_{0}$, the statistic becomes $\xi_{n}(W,Z)$. Under our independence assumption, $\sqrt{n}\xi_{n}(W,Z)\leq c_{1-\alpha}$ with probability approaching $(1-\alpha)$. Hence, our confidence set is likely to contain points around $\theta_{0}$ with probability approaching $(1-\alpha)$.

Now we proceed by verifying whether the confidence set contains regions near the boundaries of the parameter space $[0,\infty)$ (let the upper boundary be infinity).

When $\theta=0$, the statistic {becomes} $\xi_{n}(P,Z)$. If the instrument is relevant, i.e., $P\not\perp Z$, $\text{plim }\xi_{n}(P,Z)>0$ and thus $\sqrt{n}\xi_{n}(P,Z)\overset{p}{\to}\infty$. In finite samples, if $\sqrt{n}\xi_{n}(P,Z)>c_{1-\alpha}$, then those $\theta$s near $0$ are not likely to be covered by the confidence set. In that case, the confidence set will have a strictly positive lower bound.

When $\theta\to \infty$, let $f(x)=x/\theta$ which is strictly increasing in $x$ for any fixed $\theta>0$. Then by Lemma \ref{lem.monoinv}, our statistic is equal to $\xi_{n}(P/\theta-(P-W)/\theta_{0},Z)$, which is in turn equal to $\xi_{n}(P/\theta-1/Y,Z)$ by the first order condition $P-W=\theta_{0}/Y$. By Theorem 1.1 in \cite{chatterjee2020}, the probability limit of $\xi_{n}(P/\theta-1/Y,Z)$ and $\xi_{n}(-1/Y,Z)$ are $c\int\text{Var}(\mathbb{E}(1_{\{Z\geq t\}}|P/\theta-1/Y))d\mu(t)$ and $c\int\text{Var}(\mathbb{E}(1_{\{Z\geq t\}}|-1/Y))d\mu(t)$, respectively, where $c$ is some  constant and $\mu$ is the law of $Z$. If $\int\text{Var}(\mathbb{E}(1_{\{Z\geq t\}}|P/\theta-1/Y))d\mu(t)$ is continuous in $\theta$, for sufficiently large $\theta$, $\xi_{n}(P/\theta-1/Y,Z)$ and $\xi_{n}(-1/Y,Z)$ are close with probability approaching 1 (w.p.a.1) by consistency. In the meantime, noting $Y>0$ and applying Lemma \ref{lem.monoinv} again, we have $\xi_{n}(-1/Y,Z)=\xi_{n}(Y,Z)$. As long as $Z\not\perp Y$, $\sqrt{n}\xi_{n}(Y,Z)>c_{1-\alpha}$ w.p.a.1. Therefore, as $\theta$ moves away from $\theta_{0}$ to diverge to $+\infty$, the confidence set becomes less likely to cover $\theta$, implying upper boundedness of the confidence set.

Figure \ref{fig.shape} illustrates some possible shapes of the function $\Xi_{n}(\theta)\coloneqq \sqrt{n}\xi_{n}(P-\theta/Y,Z)$ on $[0,\infty)$ following the analysis above. The four combinations $l_{1}$-$l_{3}$, $l_{1}$-$l_{4}$, $l_{2}$-$l_{3}$ and $l_{2}$-$l_{4}$ depict four graphs of the function depending on the magnitude of $\sqrt{n}\xi_{n}(P,Z)$ (the intersection point of the vertical axis and $l_{1}$ or $l_{2}$) and $\sqrt{n}\xi_{n}(Y,Z)$ (the asymptote of $l_{3}$ or $l_{4}$). The confidence set contains all $\theta$ at which the function is below the horizontal $c_{1-\alpha}$ line. Specifically, when $\sqrt{n}\xi_{n}(W,Z)<c_{1-\alpha}$, the four cases are as follows:

Case 1. $\sqrt{n}\xi_{n}(P,Z)>c_{1-\alpha}$ and $\sqrt{n}\xi_{n}(Y,Z)>c_{1-\alpha}$. Function $\Xi_{n}(\theta)$ is $l_{1}$-$l_{3}$. The confidence set is a bounded interval $[A,B]$.

Case 2. $\sqrt{n}\xi_{n}(P,Z)>c_{1-\alpha}$ and $\sqrt{n}\xi_{n}(Y,Z)\leq c_{1-\alpha}$. Function $\Xi_{n}(\theta)$ is $l_{1}$-$l_{4}$. The confidence set is $[A,\infty)$.

Case 3. $\sqrt{n}\xi_{n}(P,Z)\leq c_{1-\alpha}$ and $\sqrt{n}\xi_{n}(Y,Z)>c_{1-\alpha}$. Function $\Xi_{n}(\theta)$ is $l_{2}$-$l_{3}$. The confidence set is $[0,B]$.

Case 4. $\sqrt{n}\xi_{n}(P,Z)\leq c_{1-\alpha}$ and $\sqrt{n}\xi_{n}(Y,Z)\leq c_{1-\alpha}$. Function $\Xi_{n}(\theta)$ is $l_{2}$-$l_{4}$. The confidence set is the entire parameter space $[0,\infty)$.

To summarize, a large $\sqrt{n}\xi_{n}(P,Z)$ leads to a large lower bound for the confidence interval whereas a large $\sqrt{n}\xi_{n}(Y,Z)$ results in a small upper bound.

\tikzset{every picture/.style={line width=0.75pt}}      
\begin{figure}[htbp]
\begin{center}
\begin{tikzpicture}[x=0.75pt,y=0.75pt,yscale=-1,xscale=1]

\path (150,235); 
\draw    (151,129) -- (504,129) ;
\draw   (178,88) .. controls (224.81,88.82) and (270.19,171.18) .. (317,172) ;
\draw  [dash pattern={on 3.75pt off 3pt on 7.5pt off 1.5pt}] (317,172) .. controls (363.81,172.82) and (412.72,53.29) .. (459.52,54.11) ; 
\draw   (177.13,137.8) .. controls (223.94,138.61) and (270.19,171.18) .. (317,172) ;
\draw  [dash pattern={on 3.75pt off 3pt on 7.5pt off 1.5pt}] (317,172) .. controls (363.81,172.82) and (410.93,155.38) .. (457.74,156.2) ; 
\draw    (177,43) -- (177,221) ;
\draw  [dash pattern={on 0.84pt off 2.51pt}]  (316,45) -- (316,227) ;
\draw    (150,197) -- (503,197) ;
\draw  [dash pattern={on 0.84pt off 2.51pt}]  (176,51) -- (482,51) ;
\draw  [dash pattern={on 0.84pt off 2.51pt}]  (178,156) -- (472,153) ;

\draw (301,199.4) node [anchor=north west][inner sep=0.75pt]   [align=left] {$\displaystyle \theta _{0}$};
\draw (221,85) node [anchor=north west][inner sep=0.75pt]   [align=left] {$\displaystyle l_{1}$};
\draw (248,160) node [anchor=north west][inner sep=0.75pt]    {$l_{2}$};
\draw (426,74.4) node [anchor=north west][inner sep=0.75pt]    {$l_{3}$};
\draw (397,163.4) node [anchor=north west][inner sep=0.75pt]    {$l_{4}$};
\draw (158,199.4) node [anchor=north west][inner sep=0.75pt]    {$0$};

\draw (144,116) node [anchor=north west][inner sep=0.75pt]    {$c_{1-\alpha}$};

\draw (246,110) node [anchor=north west][inner sep=0.75pt]   [align=left] {$\displaystyle A$};
\draw (358,110) node [anchor=north west][inner sep=0.75pt]   [align=left] {$\displaystyle  \begin{array}{{>{\displaystyle}l}}
B\\
\end{array}$};
\end{tikzpicture}
\caption{Function $\sqrt{n}\xi_{n}(P-\theta/Y,Z)$ and the Confidence Set}
\label{fig.shape}
\end{center}
\end{figure}

We can compute $\xi_{n}(P,Z)$ and $\xi_{n}(Y,Z)$ in our empirical applications. For instance, in the gasoline demand example, $\sqrt{n}\xi_{n}(P,Z)$ is quite large (59.74). Indeed, the lower bound of the confidence interval by only inverting $\xi$ in Table \ref{tab.ci.gas} is large; the intersection approach greatly improves the lower bound of the alternative confidence interval by 2SLS. For the upper bound, we compute $\sqrt{n}\xi_{n}(Y,Z)=-0.02$, smaller than the $0.95$th quantile of $\mathcal{N}(0,0.4)$. Hence, the $\xi$-only confidence set is likely to be unbounded from above\footnote{In Table \ref{tab.ci.gas}, the upper bound in the "$\xi$ only" row is 6 because the confidence set in that case is computed by grid search over the interval $[10^{-6},6]$.}, so the intersecting approach improves its upper bound.

\section{Monotonicity, Concavity, and Optimization} 
\label{appx.monotone}

\subsection{Monotonicity and Concavity of the (Standardized) Welfare Loss Function} 

In this section, we show that function $\sum_{k=1}^{K}x_{k}\log(1+\Delta_{k}y_{k}^{0}/x_{k})$ is component-wise strictly increasing in each $x_{k}$ on $(\max\{0,-\Delta_{k}y_{k}^{0}\},\infty)$ as long as $\Delta_{k}y_{k}^{0}\neq 0$ (and $\Delta_{k}>-\theta_{k}/y_{k}^{0}$ to guarantee the welfare loss function at the true parameter $(\theta_{1},\ldots,\theta_{k},\ldots,\theta_{K})$ is well-defined), and is strictly concave in $\bm{x}$ on $(0,\infty)^{K}$.

\subsubsection*{Monotonicity} For each $k$ and $\Delta_{k}y_{k}^{0}\neq 0$,
\begin{align*}
\frac{d}{dx_{k}}\left[x_{k}\log(1+\Delta_{k}y_{k}^{0}/x_{k})\right]=&\log\left(1+\frac{\Delta_{k}y_{k}^{0}}{x_{k}}\right)-\frac{\Delta_{k}y_{k}^{0}}{\Delta_{k}y_{k}^{0}+x_{k}}\\
>&\frac{\Delta_{k}y_{k}^{0}/x_{k}}{1+\Delta_{k}y_{k}^{0}/x_{k}}-\frac{\Delta_{k}y_{k}^{0}}{\Delta_{k}y_{k}^{0}+x_{k}}\\
=&0,
\end{align*}
where the inequality is by the fact that $\log(1+c)>c/(1+c)$ for all $c\in(-1,0)\cup(0,\infty)$\footnote{We thank an anonymous referee for suggesting the use of inequality $\log(1+c)>c/(1+c)$ to conveniently recover the entire region on which the function under consideration is strictly increasing.}. Hence, $\sum_{k=1}^{K}x_{k}\log(1+\Delta_{k}y_{k}^{0}/x_{k})$ is component-wise strictly increasing in each $x_{k}$ on $(\max\{0,-\Delta_{k}y_{k}^{0}\},\infty)$ if $\Delta_{k}y_{k}^{0}\neq 0$. 
\subsubsection*{Concavity} For each $k$,
\[
\frac{d^{2}}{dx_{k}^{2}}\left[x_{k}\log(1+\Delta_{k}y_{k}^{0}/x_{k})\right]=-\frac{(\Delta_{k}y_{k}^{0})^{2}}{x_{k}(\Delta_{k}y_{k}^{0}+x_{k})^{2}}<0,\forall x_{k}>0.
\]
Therefore, $\sum_{k=1}^{K}\theta_{k}\log(1+c_{k}/\theta_{k})$ is strictly concave on $(0,\infty)^{K}$ as long as $\Delta_{k}y_{k}^{0}\neq 0$.

\subsection{Convex Optimization}\label{appx.implement}

Since the (standardized) welfare loss function is concave in $\bm{\theta}$, we can maximize it by convex optimization algorithms when there are additional convex constraints. Many popular computer packages for convex optimization such as \texttt{CVXR} for R \citep{CVXR} and \texttt{CVXPY} for Python \citep{diamond2016cvxpy} use disciplined convex programming (DCP); DCP restricts the functions that can appear in the optimization problem and the way that the functions are composed. Specifically, such functions are called \textit{atomic functions} and have known curvatures. Examples of such functions are $-\log(x)$ and $x^{2}$. The summands of our objective function \eqref{eq.CS} are not atomic functions. In this appendix, we introduce a technique to transform our problem to be DCP-solvable. It may be of independent interest.

The goal is to transform our objective function to be the sum of atomic functions $\text{kl\_div}$, defined as $\text{kl\_div}(x_{1},x_{2})\coloneqq x_{1}\log(x_{1}/x_{2})-x_{1}+x_{2}$ for $x_{1}>0$ and $x_{2}>0$. To do so, we introduce an auxiliary variable $\bm{\gamma}\coloneqq (\gamma_{k})$ with an additional set of linear constraints $\tilde{\theta}_{k}-\gamma_{k}=-\Delta_{k}y_{k}^{0}$ for all $k$. Since we only consider positive price change, $\gamma_{k}>0$ for all $k$. Then our welfare loss function becomes
\begin{align*}
\sum_{k}\tilde{\theta}_{k}\log\left(1+\frac{\Delta_{k}y_{k}^{0}}{\tilde{\theta}_{k}}\right)=&-\sum_{k}\tilde{\theta}_{k}\log\left(\frac{\tilde{\theta}_{k}}{\tilde{\theta}_{k}+\Delta_{k}y_{k}^{0}}\right)\\
=&-\sum_{k}\left[\tilde{\theta}_{k}\log\left(\frac{\tilde{\theta}_{k}}{\tilde{\theta}_{k}+\Delta_{k}y_{k}^{0}}\right)-\tilde{\theta}_{k}+(\tilde{\theta}_{k}+\Delta_{k}y_{k}^{0})\right]+\sum_{k}\Delta_{k}y_{k}^{0}\\
=&-\sum_{k}\left[\tilde{\theta}_{k}\log\left(\frac{\tilde{\theta}_{k}}{\gamma_{k}}\right)-\tilde{\theta}_{k}+\gamma_{k}\right]+\sum_{k}\Delta_{k}y_{k}^{0}\\
=&-\sum_{k}\text{kl\_div}(\tilde{\theta}_{k},\gamma_{k})+\sum_{k}\Delta_{k}y_{k}^{0}.
\end{align*}
Since $\sum_{k}\Delta_{k}y_{k}^{0}$ does not contain $(\bm{\tilde{\theta}},\bm{\gamma})$, we can solve the following minimization problem by DCP, obtain the minimizer, and substitute it into the original objective function:
\begin{align*}
\min_{\bm{\tilde{\theta}},\bm{\gamma}}\ \ &\frac{1}{\sum_{k=1}^{K}\Delta_{k}^{2}}\sum_{k}\text{kl\_div}(\tilde{\theta}_{k},\gamma_{k}),\\
s.t.\ \ &\tilde{\bm{\theta}}\in CS(1-\alpha)\\
& \tilde{\theta}_{k}-\gamma_{k}=-\Delta_{k}y_{k}^{0},\ \forall k,\\
&f_{j}(\tilde{\theta}_{1},\ldots,\tilde{\theta}_{K})\leq 0,\ j=1,\ldots,l, \\
&g_{j}(\tilde{\theta}_{1},\ldots,\tilde{\theta}_{K})=0,\ j=1,\ldots,m,
\end{align*}
where the convex functions $f_{1},\ldots,f_{l}$ and affine functions $g_{1},\ldots,g_{m}$ characterize additional constraints on the $\tilde{\theta}_{k}$s, for instance $\sum_{k=1}^{K}\tilde{\theta}_{k}=1$.

\section{Generalization of the Utility Function} 
\label{sec:generalization_of_the_utility_function}

One key step in our method is to express the welfare change as a known function of $(\bm{y}^{0},\bm{\theta},\bm{\Delta})$ by getting rid of the unobservable vector $\bm{W}$. In this section, we show that this is not driven by the choice of the log functions in the utility function \eqref{eq.utility}. We derive a similar result under a more general utility function. 

Assume that a consumer with income level $I$ solves the following problem:
\begin{align*}
\max_{Y_{0},\bm{Y}}&\ \  U(Y_{0},\bm{Y},\bm{W},\bm{\theta})\coloneqq Y_{0}+U_{0}(\bm{Y};\bm{\theta})+\bm{W}'\bm{Y}\label{eq.utility}\\
s.t.&\ \ Y_{0}+\bm{P}' \bm{Y}=I,
\end{align*}
where $U_{0}(\cdot;\bm{\theta})$ is differentiable with the gradient denoted by $\nabla U_{0}(\cdot;\bm{\theta})$. We can solve for the optimal consumption $\bm{Y}^{*}$ as the solution to the following equation:
\begin{equation}
  \nabla U_{0}(\bm{Y}^{*};\bm{\theta})=\bm{P}-\bm{W}.\label{eq.foc.general}
\end{equation}
Suppose at the true $\bm{\theta}$, function $\nabla U_{0}(\cdot;\bm{\theta})$ is one-to-one. Denoting its inverse by $D(\cdot;\bm{\theta})$, we then have
\begin{equation}\label{eq.demand.general}
   \bm{Y}^{*}(\bm{P},\bm{W};\bm{\theta})=D(\bm{P}-\bm{W};\bm{\theta}).
 \end{equation} 
By $Y_{0}=I-\bm{P}'\bm{Y}$ and suppressing the dependence of $\bm{Y}^{*}$ on $(\bm{P},\bm{W};\bm{\theta})$, we obtain the indirect utility as follows:
\begin{align*}
  V(I,\bm{P},\bm{W},\bm{\theta})=I+U_{0}\left(\bm{Y}^{*};\bm{\theta}\right)-\left(\bm{P}-\bm{W}\right)'\bm{Y}^{*}=I+U_{0}\left(\bm{Y}^{*};\bm{\theta}\right)-\nabla U_{0}(\bm{Y}^{*};\bm{\theta})'\bm{Y}^{*},
\end{align*}
where the second equality is by equation \eqref{eq.foc.general}. 

Now same as in Section \ref{sec.welfare}, we consider the indirect utility function at price level $\bm{p}^{0}$ and $\bm{p}^{1}$ under $\bm{W}=\bar{\bm{w}}$. At $(\bm{p}^{1},\bar{\bm{w}})$, the counterfactual consumption $\bm{y}^{1}\equiv \bm{Y}^{*}(\bm{p}^{1},\bar{\bm{w}};\bm{\theta})$ satisfy the following equation:
\begin{align*}
  \bm{y}^{1}=&D\left(\bm{p}^{1}-\bar{\bm{w}};\bm{\theta}\right)=D\left(\bm{p}^{0}-\bar{\bm{w}}+\bm{\Delta};\bm{\theta}\right)
  =D\left(\nabla U_{0}\left(\bm{y}^{0};\bm{\theta}\right)+\bm{\Delta};\bm{\theta}\right),
\end{align*}
where the third equality is by $\bm{y}^{0}\equiv \bm{Y}^{*}(\bm{p}^{0},\bar{\bm{w}};\bm{\theta})$ and by equation \eqref{eq.demand.general}.
Therefore,
\begin{align*}
&\text{WL}(\bm{\theta})\\
=&V(I,\bm{p}^{0},\bar{\bm{w}},\bm{\theta})-V(I,\bm{p}^{1},\bar{\bm{w}},\bm{\theta})\\
=&U_{0}\left(\bm{y}^{0};\bm{\theta}\right)-U_{0}\left(D\left(\nabla U_{0}\left(\bm{y}^{0};\bm{\theta}\right)+\bm{\Delta};\bm{\theta}\right);\bm{\theta}\right)\\
&-\Big[\nabla U_{0}(\bm{y}^{0};\bm{\theta})'\bm{y}^{0}-\nabla U_{0}(D\left(\nabla U_{0}\left(\bm{y}^{0};\bm{\theta}\right)+\bm{\Delta};\bm{\theta}\right);\bm{\theta})'\cdot D\left(\nabla U_{0}\left(\bm{y}^{0};\bm{\theta}\right)+\bm{\Delta};\bm{\theta}\right)\Big].
\end{align*}
Although the expression is complicated, $U_{0}$, $\nabla U_{0}$ and $D$ are all known functions up to the unknown parameters $\bm{\theta}$. As $\bm{y}^{0}$ and $\bm{\Delta}$ are specified by the researcher, we can still optimize $\text{WL}(\bm{\theta})$ constrained by a confidence set of $\bm{\theta}$. In particular, one can verify that when $U_{0}(\bm{Y};\bm{\theta})=\sum_{k=1}^{K}\theta_{k}\log(Y_{k})$, the above expression for $\text{WL}(\bm{\theta})$ is equal to equation \eqref{eq.CS}.
\begin{remark}
The function $\text{WL}(\bm{\theta})$ can still be optimized efficiently when it is concave. When that is no longer the case under the general $U_{0}$, one can optimize it by other nonconvex global optimization methods. Alternatively, if we assume $U_{0}(\bm{Y};\bm{\theta})=\sum_{k=1}^{K}U_{0,k}(Y_{k};\theta_{k})$ for some $U_{0,k}$s with known functional forms, then $\text{WL}(\bm{\theta})$ will become a sum of functions; each of these functions only depends on $(y^{0}_{k},\theta_{k},\Delta_{k})$ for one $k$; the specification adopted in the main text is one such example. Although these functions may not be globally concave, one may optimize each of them by grid search within the confidence set since there is only one unknown parameter. In this way, optimization can still be done efficiently and parallelly in $k$.
\end{remark}

\section{Empirical Applications: Robustness Check}
\subsection{Gasoline Demand: Smoothing the Instrument}\label{appx.gas}
In the gasoline demand example in Section \ref{sec.gas}, the instrumental variable takes on 34 values ranging from 0.361 to 3.391. We treat this variable as continuous in the main text. In this appendix, we further smooth it by adding an independent $\text{Unif}[-0.1,0.1]$ noise term when applying Chatterjee's statistic. For 2SLS, we use the original instrument. The results are almost identical to those in Section \ref{sec.gas}; only the lower bound of the confidence interval using the intersection method and the corresponding lower bounds on the standardized welfare loss at the 0.9th, 0.5th, and 0.1th quantile change to 1.396, 5.275, 3.033, and 1.473, respectively. The intersection approach still yields the shortest intervals between the welfare loss bounds. The results are also robust to the magnitude of the uniform noise term.
\subsection{Food Demand: Impact of Prices of the Other Goods}\label{appx.food}
Under our utility function, the demand function \eqref{eq.demand} says that the prices of good $k'\neq k$ do not affect the demand for good $k$. In this section, we examine this assumption using the food demand data in Section \ref{sec.food}.

\begin{figure}[htbp]
\begin{center}
\includegraphics[scale=0.5]{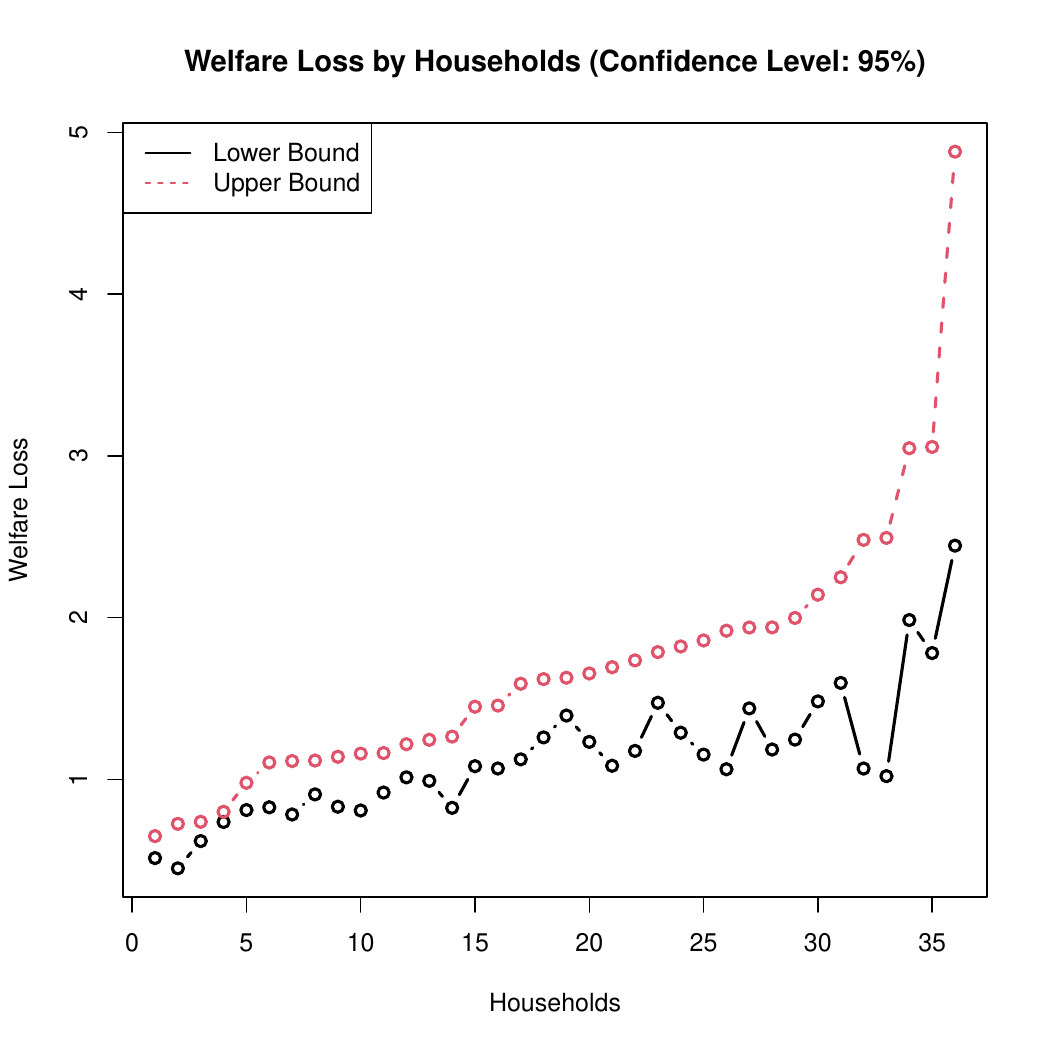}
\caption{Household-Level Standardized Welfare Loss Bounds for a Subsample (Confidence Level $=0.95$)}
\label{fig.pump.appx}
\end{center}
\end{figure}

For each household, we estimate the following system of equations by SUR:
\begin{align*}
Y_{1}^{-1}=&\alpha_{1}+\beta_{1}P_{1}+\gamma_{1,2}P_{2}+\varepsilon_{1},\\
Y_{2}^{-1}=&\alpha_{2}+\gamma_{2,1}P_{1}+\beta_{2}P_{2}+\varepsilon_{2}.
\end{align*}
We then conduct a Wald test for the hypothesis $\mathbb{H}_{0}:\gamma_{1,2}=\gamma_{2,1}=0$. At the 5\% level, this hypothesis is rejected for only 3 out of the 39 households with nonempty confidence sets under the intersection approach. We plot the welfare bounds excluding these 3 households in Figure \ref{fig.pump.appx}. We can see that it is very similar to Figure \ref{fig.pump}.

\section{Proofs}\label{sec.proof}
\subsection{Proof of Theorem \ref{thm.welfare}}

By construction, each of the three events, $\Omega_{1}\coloneqq\{\mathrm{WL}^{st}(\bm{\theta})\leq \overline{\mathrm{WL}}^{st}(1-\alpha)\}$, $\Omega_{2}\coloneqq\{\mathrm{WL}^{st}(\bm{\theta})\geq\underline{\mathrm{WL}}^{st}(1-\alpha)\}$ and $\Omega_{3}\coloneqq\{\underline{\mathrm{WL}}^{st}(1-\alpha)\leq\mathrm{WL}^{st}(\bm{\theta})\leq\overline{\mathrm{WL}}^{st}(1-\alpha)\}$, happens as long as $\bm{\theta}\in CS(1-\alpha)$. Therefore,
\begin{equation*}
\lim_{n\to\infty}\Pr\left(\Omega_{j}\right)\geq \lim_{n\to\infty}\Pr\left(\bm{\theta}\in CS(1-\alpha)\right)=1-\alpha,j=1,2,3.
\end{equation*}

\subsection{Proof of Equation \eqref{eq.2overn}}\label{sec.proof.2overn}
By i.i.d. of $\{W_{k,i}:i=1,\ldots,n\}$ and by symmetry,
\[
\underbrace{\Pr\left(W_{k,1}<W_{k,2}<\ldots <W_{k,n}\right)=\cdots=\Pr\left(W_{k,n}<W_{k,n-1}<\ldots <W_{k,1}\right)}_{n!}.
\]
Since these probabilities add up to 1 by continuity of $W_{k}$, random vector $(\pi_{k}(1),\ldots,\pi_{k}(n))$ is uniformly distributed on the set of all the $n!$ possible permutations of $\{1,\ldots,n\}$. When $|\pi_{k}(i)-\pi_{k}(j)|=1$, it is either the case that $(\pi_{k}(i),\pi_{k}(j))=((i)_{k},(i+1)_{k})$ or $((i+1)_{k},(i)_{k})$ for $(i)_{k}=(1)_{k},\ldots,(n-1)_{k}$. For each case, the other $(n-2)$ indices have $(n-2)!$ possible arrangements. Hence, in total, there are $(n-1)\times 2!\times (n-2)!$ permutations such that $|\pi_{k}(i)-\pi_{k}(j)|=1$. Therefore,
\begin{equation*}
  \Pr\left(\left|\pi_{k}(i)-\pi_{k}(j)\right|=1\right)=\frac{2(n-1)(n-2)!}{n!}=\frac{2}{n}.
\end{equation*}

\subsection{Proof of Proposition \ref{prop.suff}}
By equation \eqref{eq.2overn}, $\Pr(|\pi_{k}(i)-\pi_{k}(j)|=1)=2/n$. It is then sufficient to show that the conditional probability $\Pr\left(|\pi_{k'}(i)-\pi_{k'}(j)|=1\Big| |\pi_{k}(i)-\pi_{k}(j)|=1\right)\to 0$ for all $1\leq i<j\leq n$. Under the sure event that there are no ties in $W_{k',i}$s, $\pi_{k'}(i)$ is the rank of $W_{k,i}$, i.e., $\sum_{i'}1(W_{k',i'}\leq W_{k',i})$. We can then approximate it by $nF_{W_{k'}}(W_{k',i})$. Specifically, by \cite{smirnov1944approximate} or \cite{chung1949estimate}, under Assumptions \ref{assu.iid} and \ref{assu.cont}-i), we have the following for all $i$:
\begin{equation}\label{eq.gc}
  \left|\frac{1}{n}\sum_{i'}1(W_{k',i'}\leq W_{k',i})-F_{W_{k'}}(W_{k',i})\right|\leq\sup_{w\in\mathbb{R}}\left|\frac{1}{n}\sum_{i'}1(W_{k',i'}\leq w)-F_{W_{k'}}(w)\right|=O\left(\sqrt{\frac{\log\log{n}}{n}}\right)\ a.s.
\end{equation}
So, there exists a constant $C$ such that for all $i\neq j$,
\begin{align}
  &\Pr\left(|\pi_{k'}(i)-\pi_{k'}(j)|=1\Big| |\pi_{k}(i)-\pi_{k}(j)|=1\right)\notag\\
  =&\Pr\left(\Big|\frac{\pi_{k'}(i)}{n}-\frac{\pi_{k'}(j)}{n}\Big|=\frac{1}{n}\Bigg| |\pi_{k}(i)-\pi_{k}(j)|=1\right)\notag\\\
  \leq &\Pr\left(\Big|F_{W_{k'}}(W_{k',i})-F_{W_{k'}}(W_{k',j})\Big|\leq C\sqrt{\frac{\log\log{n}}{n}}\Bigg| |\pi_{k}(i)-\pi_{k}(j)|=1\right)\notag\\\
  =&\mathbb{E}\left[\Pr\left(\Big|F_{W_{k'}}(W_{k',i})-F_{W_{k'}}(W_{k',j})\Big|\leq C\sqrt{\frac{\log\log{n}}{n}}\Bigg|W_{k',j},W_{k,1},\ldots,W_{k,n}\right)\Bigg| |\pi_{k}(i)-\pi_{k}(j)|=1\right],\label{eq.interm}
\end{align}
where the inequality is by equation \eqref{eq.gc} and the triangle inequality. The last equality is by the law of iterated expectation by noticing that $\pi_{k}$ is a function of $W_{k,1},\ldots,W_{k,n}$.
Now we consider the conditional probability inside. For every realization $w$ of $W_{k',j}$ and $v'$ such that $F_{W_{k',j}}(w)=v'$,
\begin{align*}
&\Pr\left(\Big|F_{W_{k'}}(W_{k',i})-F_{W_{k'}}(W_{k',j})\Big|\leq C\sqrt{\frac{\log\log{n}}{n}}\Bigg| W_{k',j}=F^{-1}_{W_{k',j}}(v'),W_{k,1},\ldots,W_{k,n}\right)\\
=&\Pr\left(V_{k',i}\in \left[v'-C\sqrt{\frac{\log\log{n}}{n}},v'+ C\sqrt{\frac{\log\log{n}}{n}}\right]\Bigg| W_{k,i}\right),
\end{align*}
where the conditioning variables except for $W_{k,i}$ are dropped because $W_{k',i}$ is independent of them conditional on $W_{k,i}$ by Assumption \ref{assu.iid}. 
By condition \eqref{eq.suff}, the conditional probability converges to 0 almost surely. Therefore, by the dominated convergence theorem, the conditional expectation on the right-hand side of the last equality of \eqref{eq.interm} converges to 0.
\subsection{Proof of Theorem \ref{thm.joint1}}
Throughout, assume that there are no ties in the sample $\{Z_{ki}:i=1\ldots n\}$ for all $k$. This event happens with probability 1 by continuity of $(Z_{1},...,Z_{K})$. We first prove the following lemmas.

\begin{lem}[Haj\'ek Representation]\label{lem.hajek}
Let $U_{k,(i)_{k}}\coloneqq F_{Z_{k}}(Z_{k,(i)_{k}})$.
Under Assumptions \ref{assu.iid} and \ref{assu.cont}, for each $k=1,\ldots,K$, 
$$\sqrt{n}\xi_{n}^{(k)}=-\frac{3}{\sqrt{n-1}}\sum_{i=1}^{n-1}\left(\left|U_{k,(i+1)_{k}}-U_{k,(i)_{k}}\right|+2U_{k,(i)_{k}}\left(1-U_{k,(i)_{k}}\right)-\frac{2}{3}\right)+o_{p}(1).$$
\end{lem} 
\begin{proof}
We first note that 
\begin{align*}
\sqrt{n}\xi_{n}(W_{k},Z_{k})\coloneqq& \sqrt{n}-\frac{3\sqrt{n}\sum_{i=1}^{n-1}|r_{k,i+1}-r_{k,i}|}{n^{2}-1}\notag\\\
=&-3\left[\frac{\sum_{i=1}^{n-1}|r_{k,i+1}-r_{k,i}|-n(n-1)/3}{\sqrt{n}(n-1)}-\frac{\sum_{i=1}^{n-1}|r_{k,i+1}-r_{k,i}|}{\sqrt{n}(n+1)(n-1)}\right]\notag\\\
=&-3\underbrace{\frac{\sum_{i=1}^{n-1}|r_{k,i+1}-r_{k,i}|-n(n-1)/3}{\sqrt{n}(n-1)}}_{A_{k}}+o_{p}(1),\label{eq.hajek.rewrite}
\end{align*}
where the last equality holds in view of $\sum_{i=1}^{n-1}|r_{k,i+1}-r_{k,i}|\leq (n-1)^{2}$. 

By construction $r_{k,i}\coloneqq \sum_{j=1}^{n}1(Z_{k,j}\leq Z_{k,(i)_{k}})$. Under the almost sure event that there no ties among $Z_{k,i}$s, $(r_{k,1},\ldots,r_{k,n})$ forms a permutation of indices $1,\ldots,n$. By $Z_{k}\perp W_{k}$ and because $Z_{k,i}$s are i.i.d. across $i$, we have that $U_{k,(i)_{k}}\sim\text{Unif}(0,1)$ are i.i.d. across $i$, and that the random permutation $(r_{k,1},\ldots,r_{k,n})$ is uniformly distributed on the set of all $n!$ permutations of $1,\ldots,n$. 
By \cite{angus1995coupling} (see also \cite{shi2022power} and \cite{zhang2023asymptotic}), we thus have 
\begin{equation*}
  A_{k}=\frac{1}{\sqrt{n-1}}\sum_{i=1}^{n-1}\left(\left|U_{k,(i+1)_{k}}-U_{k,(i)_{k}}\right|+2U_{k,(i)_{k}}\left(1-U_{k,(i)_{k}}\right)-\frac{2}{3}\right)+o_{p}(1).
\end{equation*}
The desired result follows.
\end{proof}

\sloppy For an arbitrary $\varepsilon>0$, define random variable $D_{k,k'}(\varepsilon)=1$ if $\sum_{i=1}^{n-1}1\left(\left|\pi_{k'}(\pi_{k}^{-1}((i)_{k}))-\pi_{k'}(\pi_{k}^{-1}((i+1)_{k}))\right|=1\right)\leq \varepsilon (n-1)$, and $D_{k,k'}(\varepsilon)=0$ otherwise. In words, $D_{k,k'}(\varepsilon)=1$ if and only if there are at most $\varepsilon(n-1)$ pairs of original indices $(j,l)$ that are adjacent both after arranging $W_{k,i}$s in ascending order and after arranging $W_{k',i}$s in ascending order.
\begin{lem}\label{lem.de}
Under Assumptions \ref{assu.iid} to \ref{assum.dependence}, $\Pr(D_{k,k'}(\varepsilon)=1)\to 1$ for all $\varepsilon>0$ and all $k\neq k'$.
\end{lem}
\begin{proof}

Let $Q_{(i)}\coloneqq 1\left(|\pi_{k'}(\pi_{k}^{-1}((i+1)_{k}))-\pi_{k'}\left(\pi_{k}^{-1}((i)_{k})\right)|=1\right)$.
For all $k=1,\ldots,K$, denote the vector $(W_{k,1},\ldots,W_{k,n})$ by $\bm{W}_{k}$. First, note that $\sum_{i=1}^{n-1}Q_{(i)}$ is a function of $(\bm{W}_{k},\bm{W}_{k'})$ which is well defined even if there are ties in $\bm{W}_{k}$ or $\bm{W}_{k'}$ because $(\pi_{k}(i))$ and $(\pi_{k'}(i))$ are always permutations of $\{1,\ldots,n\}$ by definition. Consider an arbitrary realization of $(\bm{W}_{k},\bm{W}_{k'})=(\bm{w}_{k},\bm{w}_{k'})$.  For any fixed $i$, suppose we replace $(w_{k,i},w_{k',i})$ by another value $(w_{k,i}',w_{k',i}')$. Denote the realization after the replacement by $(\bm{w}_{k}',\bm{w}_{k'}')$. After the replacement, index $i$ could be removed from the original slot in the permutation $(\pi_{k}(1),\ldots,\pi_{k}(n))$ and/or $(\pi_{k'}(1),\ldots,\pi_{k'}(n))$, and be inserted into a new slot. Noticing that removing $i$ from the original slots under $(\bm{w}_{k},\bm{w}_{k'})$ may at most increase (or decrease) the number of pairs of indices that are adjacent in both permutations by 2 (or 1). Meanwhile, plugging $i$ into new slots under $(\bm{w}_{k}',\bm{w}_{k'}')$ may at most increase (or decrease) the number of pairs of indices that are adjacent in both permutations by 1 (or 2). Therefore,  $\sum_{i=1}^{n-1}Q_{(i)}$ changes at most by $3$. For example, let $n=12$ and under some realization of $\bm{W}_{k}$ and $\bm{W}_{k'}$, assume $(\pi_{k}(1),\ldots,\pi_{k}(12))=(1,3,5,7,9,11,2,4,6,8,10,12)$ and $(\pi_{k'}(1),\ldots,\pi_{k'}(12))=(1,3,5,7,9,11,2,4,6,8,12,10)$. Then $\sum_{i=1}^{n-1}Q_{(i)}=10$. Now suppose we change the realizations of $W_{k,5}$ and $W_{k',5}$ such that the new permutations are
\begin{align*}
  &(1,3,7,9,5,11,2,4,6,8,10,12),\\
  &(1,3,7,9,11,2,4,6,5,8,12,10).
\end{align*}
We can see that removing $5$ from the original third slot leads to a reduction of one pair of indices that are adjacent in both permutations; originally there were $(3,5)$ and $(5,7)$ but now only $(3,7)$. Meanwhile, inserting 5 into the new slots makes the pairs $(9,11)$ and $(6,8)$ no longer adjacent in both permutations. So function $\sum_{i=1}^{n-1}Q_{(i)}$ becomes 7 after changing the realizations of $W_{k,5}$ and $W_{k',5}$. 

Therefore, by the bounded difference concentration inequality \citep{McDiarmid_1989}, we have that for all $0<\varepsilon_{1}<\varepsilon$,
\begin{align*}
  &\Pr\left(\frac{\sum_{i=1}^{n-1}\left(Q_{(i)}-\mathbb{E}Q_{(i)}\right)}{n-1}>\varepsilon_{1}\right)\leq \exp\left(-\frac{2}{9}(n-1)\varepsilon_{1}^{2}\right).\\
  \Leftrightarrow&\Pr\left(\sum_{i=1}^{n-1}Q_{(i)}>(n-1)\varepsilon_{1}+\sum_{i=1}^{n-1}\mathbb{E}Q_{(i)}\right)\leq \exp\left(-\frac{2}{9}(n-1)\varepsilon_{1}^{2}\right).
\end{align*}
Now it suffices to show that $\sum_{i=1}^{n-1}\mathbb{E}Q_{(i)}<(n-1)(\varepsilon-\varepsilon_{1})$ for any $0<\varepsilon_{1}<\varepsilon$ for large enough $n$  because then we will have
\begin{align*}
  \Pr\left(\sum_{i=1}^{n-1}Q_{(i)}>(n-1)\varepsilon\right)\leq\Pr\left(\sum_{i=1}^{n-1}Q_{(i)}>(n-1)\varepsilon_{1}+\sum_{i=1}^{n-1}\mathbb{E}Q_{(i)}\right)\leq \exp\left(-\frac{2}{9}(n-1)\varepsilon_{1}^{2}\right)\to 0.
\end{align*}

Noticing that
\[\sum_{i=1}^{n-1}Q_{(i)}=\sum_{i=1}^{n-1}\sum_{j=i+1}^{n}1\left(|\pi_{k}(i)-\pi_{k}(j)|=1\right)\cdot 1\left(|\pi_{k'}(i)-\pi_{k'}(j)|=1\right).\]
Therefore, by Assumption \ref{assum.dependence},
\[
  \frac{\sum_{i=1}^{n-1}\mathbb{E}Q_{(i)}}{n-1}=\frac{\sum_{i=1}^{n-1}\sum_{j=i+1}^{n}\Pr\left(|\pi_{k}(i)-\pi_{k}(j)|=1\cap|\pi_{k'}(i)-\pi_{k'}(j)|=1\right)}{n-1}=o(1).
\]
The desired result follows as $k$ and $k'$ are arbitrary.
\end{proof}

\begin{lem}\label{lem.zero}
Let $U_{1},U_{1}',U_{2},U_{3}$ be distributed as $\text{Unif}(0,1)$. If $(U_{1},U_{1}')\perp (U_{2},U_{3})$ and $U_{2}\perp U_{3}$, then
\begin{align*}
&cov\left(|U_{1}-U_{2}|,|U_{1}'-U_{3}|\right)+cov\left(U_{1}(1-U_{1}),|U_{1}'-U_{3}|\right)\\
  &+cov\left(U_{1}'(1-U_{1}'),|U_{1}-U_{2}|\right)+cov\left(U_{1}(1-U_{1}),U_{1}'(1-U_{1}')\right)=0.
\end{align*}
\end{lem}
\begin{proof}
First,
\begin{align}
  &\mathbb{E}\left(|U_{1}-U_{2}|\right)\notag\\
  =&\mathbb{E}\left[\mathbb{E}\left(|U_{1}-U_{2}|\Big|U_{2}\right)\right]\notag\\
  =&\mathbb{E}\left[\Pr\left(U_{1}>U_{2}|U_{2}\right)\left(\mathbb{E}(U_{1}|U_{1}>U_{2},U_{2})-U_{2}\right)\right]+\mathbb{E}\left[\Pr\left(U_{1}\leq U_{2}|U_{2}\right)\left(U_{2}-\mathbb{E}(U_{1}|U_{1}\leq U_{2},U_{2})\right)\right]\notag\\
=&\mathbb{E}\left[\frac{(1-U_{2})(1+U_{2})}{2}-(1-U_{2})U_{2}-\frac{U_{2}^{2}}{2}+U_{2}^{2}\right]\notag\\
=&\mathbb{E}\left[\frac{1}{2}+U_{2}^{2}-U_{2}\right]\notag\\
=&\frac{1}{3},\label{eq.zero1}
\end{align}
where the third equality is by independence between $U_{1}$ and $U_{2}$.
Similarly, $\mathbb{E}\left(|U_{1}'-U_{3}|\right)=1/3$ as well. Meanwhile, 
\begin{equation}
  \mathbb{E}[U_{1}(1-U_{1})]= \mathbb{E}[U_{1}'(1-U_{1}')]=\frac{1}{6}.\label{eq.zero2}
\end{equation}

Now we consider the following expectations. First,
\begin{align}
\mathbb{E}\left[|U_{1}-U_{2}|\cdot |U_{1}'-U_{3}| \right]=&\mathbb{E}\left\{\mathbb{E}\left[\left|U_{1}-U_{2}\right|\cdot \left|U_{1}'-U_{3}\right| \Big|U_{1},U_{1}'\right]\right\}\notag\\
=&\mathbb{E}\left\{\mathbb{E}\left[\left|U_{1}-U_{2}\right|\Big|U_{1},U_{1}'\right]\cdot\mathbb{E}\left[ \left|U_{1}'-U_{3}\right| \Big|U_{1},U_{1}'\right]\right\}\notag\\
=&\mathbb{E}\left\{\mathbb{E}\left[\left|U_{1}-U_{2}\right|\Big|U_{1}\right]\cdot\mathbb{E}\left[ \left|U_{1}'-U_{3}\right| \Big|U_{1}'\right]\right\}\notag\\
=&\mathbb{E}\left\{\left(\frac{1}{2}+U_{1}^{2}-U_{1}\right)\cdot\left(\frac{1}{2}+U_{1}^{'2}-U_{1}'\right)\right\}\notag\\
=& \mathbb{E}\left[U_{1}^{2}U_{1}^{'2}+U_{1}U_{1}'-U_{1}U_{1}^{'2}-U_{1}^{2}U_{1}'\right]+\frac{1}{12},\label{eq.zero3}
\end{align}
where the first equality is by the law of iterated expectation. The second equality is because conditional on $U_{1}$ and $U_{1}'$, random variables $|U_{1}-U_{2}|$ and $|U_{1}'-U_{3}|$ are independent. The third equality is by $U_{2}\perp U_{1'}|U_{1}$ and $U_{3}\perp U_{1}|U_{1}'$. The fourth equality follows the derivation of equation \eqref{eq.zero1}.

Next, following a similar argument, we have
\begin{align}
\mathbb{E}\left[U_{1}(1-U_{1})\cdot |U_{1}'-U_{3}| \right]=&\mathbb{E}\left\{U_{1}(1-U_{1})\mathbb{E}\left[\left|U_{1}'-U_{3}\right| \Big|U_{1},U_{1}'\right]\right\}\notag\\
=&\mathbb{E}\left\{U_{1}(1-U_{1})\left(\frac{1}{2}+U_{1}^{'2}-U_{1}'\right)\right\}\notag\\
=&\mathbb{E}\left(-U_{1}^{2}U_{1}^{'2}-U_{1}U_{1}'+U_{1}U_{1}^{'2}+U_{1}^{2}U_{1}'\right)+\frac{1}{12},\label{eq.zero4}
\end{align}
and
\begin{align}
\mathbb{E}\left[U_{1}'(1-U_{1}')|\cdot |U_{1}-U_{2}| \right]=\mathbb{E}\left(-U_{1}^{2}U_{1}^{'2}-U_{1}U_{1}'+U_{1}U_{1}^{'2}+U_{1}^{2}U_{1}'\right)+\frac{1}{12}.\label{eq.zero5}
\end{align}

Finally,
\begin{align}
\mathbb{E}\left[U_{1}(1-U_{1})\cdot U_{1}'(1-U_{1}') \right]=\mathbb{E}\left(U_{1}^{2}U_{1}^{'2}+U_{1}U_{1}'-U_{1}U_{1}^{'2}-U_{1}^{2}U_{1}'\right).\label{eq.zero6}
\end{align}
Combining equations \eqref{eq.zero1} to \eqref{eq.zero6} yields
\begin{align*}
&cov\left(|U_{1}-U_{2}|,|U_{1}'-U_{3}|\right)+cov\left(U_{1}(1-U_{1}),|U_{1}'-U_{3}|\right)\\
  &+cov\left(U_{1}'(1-U_{1}'),|U_{1}-U_{2}|\right)+cov\left(U_{1}(1-U_{1}),U_{1}'(1-U_{1}')\right)\\
  =&\frac{1}{12}+\frac{1}{12}+\frac{1}{12}-\frac{1}{9}-\frac{1}{18}-\frac{1}{18}-\frac{1}{36}\\
  =&0.
\end{align*}
\end{proof}
\begin{lem}\label{lem.cov}
Let $A_{k,(i)_{k}}\coloneqq \left|U_{k,(i+1)_{k}}-U_{k,(i)_{k}}\right|+2U_{k,(i)_{k}}(1-U_{k,(i)_{k}})-2/3$. Under Assumptions \ref{assu.iid} to \ref{assum.dependence}, $cov\left(\sum_{i=1}^{n-1}A_{k,(i)_{k}},\sum_{i=1}^{n-1}A_{k',(i)_{k'}}\right)/n=o(1)$ for all $k\neq k'$.
\end{lem}
\begin{proof}
For any $0<\varepsilon<1$, by the law of total covariance, we have the following:
\begin{align}
&cov\left(\sum_{i=1}^{n-1}A_{k,(i)_{k}},\sum_{i=1}^{n-1}A_{k',(i)_{k'}}\right)\notag\\
=&\mathbb{E}\left[cov\left(\sum_{i=1}^{n-1}A_{k,(i)_{k}},\sum_{i=1}^{n-1}A_{k',(i)_{k'}}\Bigg|D_{k,k'}(\varepsilon)\right)\right]+cov\left[\mathbb{E}\left(\sum_{i=1}^{n-1}A_{k,(i)_{k}}\Bigg|D_{k,k'}(\varepsilon)\right),\mathbb{E}\left(\sum_{i=1}^{n-1}A_{k',(i)_{k'}}\Bigg|D_{k,k'}(\varepsilon)\right)\right].\label{eq.cov1}
\end{align}
Let $\bm{W}_{k}=(W_{k,1},\ldots,W_{k,n})$ for all $k$. Applying the law of total covariance again to the first term on the right-hand side of equation \eqref{eq.cov1},
\begin{align}
&\mathbb{E}\left[cov\left(\sum_{i=1}^{n-1}A_{k,(i)_{k}},\sum_{i=1}^{n-1}A_{k',(i)_{k'}}\Bigg|D_{k,k'}(\varepsilon)\right)\right]\notag\\
=&\mathbb{E}\left\{\mathbb{E}\left[cov\left(\sum_{i=1}^{n-1}A_{k,(i)_{k}},\sum_{i=1}^{n-1}A_{k',(i)_{k'}}\Bigg|D_{k,k'}(\varepsilon),\bm{W}_{k},\bm{W}_{k'}\right)\Bigg|D_{k,k'}(\varepsilon)\right]\right\}\notag\\
&+\mathbb{E}\left\{cov\left[\mathbb{E}\left(\sum_{i=1}^{n-1}A_{k,(i)_{k}}\Bigg|D_{k,k'}(\varepsilon),\bm{W}_{k},\bm{W}_{k'}\right),\mathbb{E}\left(\sum_{i=1}^{n-1}A_{k',(i)_{k'}}\Bigg|D_{k,k'}(\varepsilon),\bm{W}_{k},\bm{W}_{k'}\right)\Bigg|D_{k,k'}(\varepsilon)\right]\right\}.\label{eq.cov2}
\end{align}

For the second term on the right-hand side of equation \eqref{eq.cov1}, the law of iterated expectation implies that
\begin{align}
cov\Bigg[\mathbb{E}&\left(\sum_{i=1}^{n-1}A_{k,(i)_{k}}\Bigg|D_{k,k'}(\varepsilon)\right),\mathbb{E}\left(\sum_{i=1}^{n-1}A_{k',(i)_{k'}}\Bigg|D_{k,k'}(\varepsilon)\right)\Bigg]\notag\\
=cov\Bigg[\mathbb{E}&\left\{\mathbb{E}\left(\sum_{i=1}^{n-1}A_{k,(i)_{k}}\Bigg|D_{k,k'}(\varepsilon),\bm{W}_{k},\bm{W}_{k'}\right)\Bigg|D_{k,k'}(\varepsilon)\right\},\notag\\
\mathbb{E}&\left\{\mathbb{E}\left(\sum_{i=1}^{n-1}A_{k',(i)_{k'}}\Bigg|D_{k,k'}(\varepsilon),\bm{W}_{k},\bm{W}_{k'}\right)\Bigg|D_{k,k'}(\varepsilon)\right\}\Bigg].\label{eq.cov3}
\end{align}

By independence between $(W_{1},\ldots,W_{K})$ and $(Z_{1},\ldots,Z_{K})$ and i.i.d, equations \eqref{eq.zero1} and \eqref{eq.zero2} imply that $\mathbb{E}\left(\sum_{i=1}^{n-1}A_{k,(i)_{k}}\Bigg|D_{k,k'}(\varepsilon),\bm{W}_{k},\bm{W}_{k'}\right)=0$ for all realizations of $D_{k,k'}(\varepsilon),\bm{W}_{k}$ and $\bm{W}_{k'}$. Therefore, the right-hand side of equation \eqref{eq.cov3} and thus the second term on the right-hand side of \eqref{eq.cov1} are zero. Similarly, the second term on the right-hand side of equation \eqref{eq.cov2} is also zero.  Substituting them into equation \eqref{eq.cov1},
\begin{align}
  &cov\left(\sum_{i=1}^{n-1}A_{k,(i)_{k}},\sum_{i=1}^{n-1}A_{k',(i)_{k'}}\right)\notag\\
  =&\mathbb{E}\left\{\mathbb{E}\left[cov\left(\sum_{i=1}^{n-1}A_{k,(i)_{k}},\sum_{i=1}^{n-1}A_{k',(i)_{k'}}\Bigg|D_{k,k'}(\varepsilon),\bm{W}_{k},\bm{W}_{k'}\right)\Bigg|D_{k,k'}(\varepsilon)\right]\right\}\notag\\
  =&\Pr\left(D_{k,k'}(\varepsilon)=1\right)\mathbb{E}\left[cov\left(\sum_{i=1}^{n-1}A_{k,(i)_{k}},\sum_{i=1}^{n-1}A_{k',(i)_{k'}}\Bigg|D_{k,k'}(\varepsilon)=1,\bm{W}_{k},\bm{W}_{k'}\right)\Bigg|D_{k,k'}(\varepsilon)=1\right]\notag\\
  &+\Pr\left(D_{k,k'}(\varepsilon)=0\right)\mathbb{E}\left[cov\left(\sum_{i=1}^{n-1}A_{k,(i)_{k}},\sum_{i=1}^{n-1}A_{k',(i)_{k'}}\Bigg|D_{k,k'}(\varepsilon)=0,\bm{W}_{k},\bm{W}_{k'}\right)\Bigg|D_{k,k'}(\varepsilon)=0\right].\label{eq.cov4}
\end{align}

Again, by independence between $(W_{1},\ldots,W_{K})$ and $(Z_{1},\ldots,Z_{K})$, by i.i.d., and by boundedness of $U_{k}$ and $U_{k'}$, there exists a constant $C_{1}$ such that 
\begin{equation*}
  \left|cov\left(\sum_{i=1}^{n-1}A_{k,(i)_{k}},\sum_{i=1}^{n-1}A_{k',(i)_{k'}}\Bigg|D_{k,k'}(\varepsilon)=0,\bm{W}_{k},\bm{W}_{k'}\right)\right|\leq nC_{1},a.s.
\end{equation*}
Therefore, 
\begin{equation}\label{eq.cov4.5}
\frac{1}{n}\left|\mathbb{E}\left[cov\left(\sum_{i=1}^{n-1}A_{k,(i)_{k}},\sum_{i=1}^{n-1}A_{k',(i)_{k'}}\Bigg|D_{k,k'}(\varepsilon)=0,\bm{W}_{k},\bm{W}_{k'}\right)\Bigg|D_{k,k'}(\varepsilon)=0\right]\right|\leq C_{1}.
\end{equation}

Now we consider the first term on the right-hand side of equation \eqref{eq.cov4}. Let set $\mathcal{I}\coloneqq \{(i)_{k}:|\pi_{k'}(\pi_{k}^{-1}((i+1)_{k}))-\pi_{k'}(\pi_{k}^{-1}((i)_{k}))|=1\}$. Under $D_{k,k'}(\varepsilon)=1$, $|\mathcal{I}|\leq \varepsilon(n-1)$. Let $U_{k,i}=F_{Z_{k}}(Z_{k,i})$ and $U_{k',i}=F_{Z_{k'}}(Z_{k',i})$. By i.i.d. and by independence between $(W_{1},\ldots,W_{K})$ and $(Z_{1},\ldots,Z_{K})$, once conditional on $(\bm{W}_{k},\bm{W}_{k'})$, $U_{k,(i)_{k}}$ and $U_{k',(i)_{k'}}$ are i.i.d. in $(i)_{k}$ and $(i)_{k'}$ respectively. Therefore, by the definition of $A_{k,(i)_{k}}$ and $A_{k',(i)_{k'}}$
\begin{align}
&cov\left(\sum_{i=1}^{n-1}A_{k,(i)_{k}},\sum_{i=1}^{n-1}A_{k',(i)_{k'}}\Bigg|D_{k,k'}(\varepsilon)=1,\bm{W}_{k},\bm{W}_{k'}\right)\notag\\
=&\underbrace{cov\left(\sum_{(i)_{k}\in\mathcal{I}}\left|U_{k,(i+1)_{k}}-U_{k,(i)_{k}}\right|,\sum_{i=1}^{n-1}\left|U_{k',(i+1)_{k'}}-U_{k',(i)_{k'}}\right|\Big|D_{k,k'}(\varepsilon)=1,\bm{W}_{k},\bm{W}_{k'}\right)}_{M_{1}}\notag\\
&+\underbrace{cov\left(\sum_{(i)_{k}\notin\mathcal{I}}\left|U_{k,(i+1)_{k}}-U_{k,(i)_{k}}\right|,\sum_{i=1}^{n-1}\left|U_{k',(i+1)_{k'}}-U_{k',(i)_{k'}}\right|\Big|D_{k,k'}(\varepsilon)=1,\bm{W}_{k},\bm{W}_{k'}\right)}_{M_{2}}\notag\\
&+\underbrace{2cov\left(\sum_{(i,j):\pi_{k}(i)=(1)_{k},\pi_{k}(j)=(2)_{k}}^{{(i,j):\pi_{k}(i)=(n-1)_{k},\pi_{k}(j)=(n)_{k}}}\left|U_{k,j}-U_{k,i}\right|,\sum_{i=1}^{n}U_{k',i}\left(1-U_{k',i}\right)\Big|D_{k,k'}(\varepsilon)=1,\bm{W}_{k},\bm{W}_{k'}\right)}_{M_{3}}\notag\\
&+\underbrace{2cov\left(\sum_{(i,j):\pi_{k'}(i)=(1)_{k'},\pi_{k'}(j)=(2)_{k'}}^{{(i,j):\pi_{k'}(i)=(n-1)_{k'},\pi_{k'}(j)=(n)_{k'}}}\left|U_{k',j}-U_{k',i}\right|,\sum_{i=1}^{n}U_{k,i}\left(1-U_{k,i}\right)\Big|D_{k,k'}(\varepsilon)=1,\bm{W}_{k},\bm{W}_{k'}\right)}_{M_{4}}\notag\\
&+\underbrace{4cov\left(\sum_{i=1}^{n}U_{k,i}\left(1-U_{k,i}\right),\sum_{i=1}^{n}U_{k',i}\left(1-U_{k',i}\right)\right)}_{M_{5}}+O(1),\label{eq.cov5}
\end{align}
where in $M_{3}$, $M_{4}$ and $M_{5}$, $\sum_{i=1}^{n-1}U_{k,(i)_{k}}(1-U_{k,(i)_{k}})$ and $\sum_{i=1}^{n-1}U_{k',(i)_{k'}}(1-U_{k',(i)_{k'}})$ are replaced by $\sum_{i=1}^{n}U_{k,i}(1-U_{k,i})$ and $\sum_{i=1}^{n}U_{k',i}(1-U_{k',i})$, respectively. Such replacement introduces one extra term into the summation because for any realization of $\bm{W}_{k}$ and $\bm{W}_{k'}$,  $\sum_{i=1}^{n-1}U_{k,(i)_{k}}(1-U_{k,(i)_{k}})$ is the sum of $U_{k,i}(1-U_{k,i})$ over $i$ in a subset of $\{1,\ldots,n\}$ that has $(n-1)$ elements. Since the index corresponding to that extra term only shows up at most twice in the other summation in the covariances considered, the differences in the covariances before and after the replacement are bounded by some constant by i.i.d. Hence we have an $O(1)$ term at the end of the right-hand side of equation \eqref{eq.cov5}. Now we consider $M_{1}$ to $M_{5}$.

$M_{1}$. Under $D_{k,k'}(\varepsilon)=1$, we have $|\mathcal{I}|\leq \varepsilon(n-1)$. By the boundedness of $U_{k}$ and $U_{k'}$, by i.i.d. and by independence between $(W_{1},\ldots,W_{K})$ and $(Z_{1},\ldots,Z_{K})$, there exists some $C_{2}>0$ which does not depend on $\varepsilon$ such that 
\begin{equation}\label{eq.covM1}
|M_{1}|\leq C_{2}\varepsilon(n-1).
\end{equation}

$M_{2}$. If neither $\pi_{k'}(\pi_{k}^{-1}((i)_{k}))$ nor $\pi_{k'}(\pi_{k}^{-1}((i+1)_{k}))$ is $(1)_{k'}$ or $(n)_{k'}$, by $(i)_{k}\notin\mathcal{I}$, indices $\pi_{k}^{-1}((i)_{k})$ and $\pi_{k}^{-1}((i+1)_{k})$ will show up in four different terms among the summands of $\sum_{i=1}^{n-1}\left|U_{k',(i+1)_{k'}}-U_{k',(i)_{k'}}\right|$. To see this, for an arbitrary realization of $\bm{W}_{k}$ and $\bm{W}_{k'}$, let $\pi_{k}(l)=(i)_{k}$ and $\pi_{k}(m)=(i+1)_{k}$. Let $\pi_{k'}(l)=(j)_{k'}$ where $1<j<n$. Let $\pi_{k'}^{-1}((j+1)_{k'})=l_{1}$ and $\pi_{k'}^{-1}((j-1)_{k'})=l_{2}$.  By $(i)_{k}\not\in \mathcal{I}$, we have $m\neq l_{1}\neq l_{2}$. Similarly, let $m_{1}=\pi_{k'}^{-1}(\pi_{k'}(m)+1)$ and $m_{2}=\pi_{k'}^{-1}(\pi_{k'}(m)-1)$ as $1<\pi_{k'}(m)<n$. Then $l\neq m_{1}\neq m_{2}$. Therefore, in this case, we have
\begin{align*}
&cov\left(\left|U_{k,(i+1)_{k}}-U_{k,(i)_{k}}\right|,\sum_{i=1}^{n-1}\left|U_{k',(i+1)_{k'}}-U_{k',(i)_{k'}}\right|\Big|D_{k,k'}(\varepsilon)=1,\bm{W}_{k},\bm{W}_{k'}\right)\\
=&cov\left(\left|U_{k,m}-U_{k,l}\right|,\left|U_{k',m_{1}}-U_{k',m}\right|\right)+cov\left(\left|U_{k,m}-U_{k,l}\right|,\left|U_{k',m}-U_{k',m_{2}}\right|\right)\\
&+cov\left(\left|U_{k,m}-U_{k,l}\right|,\left|U_{k',l_{1}}-U_{k',l}\right|\right)+cov\left(\left|U_{k,m}-U_{k,l}\right|,\left|U_{k',l}-U_{k',l_{2}}\right|\right)\\
=&4cov\left(\left|U_{k,1}-U_{k,2}\right|,\left|U_{k',1}-U_{k',3}\right|\right),
\end{align*}
where the first equality is by $(i)_{k}\not\in\mathcal{I}$ and by independence between $(W_{1},\ldots,W_{K})$ and $(Z_{1},\ldots,Z_{K})$. The last equality is by i.i.d. Therefore, noting that $|\mathcal{I}^{c}|\geq (1-\varepsilon)(n-1)$ conditional on $D_{k,k'}(\varepsilon)=1$,
\begin{align}
M_{2}=4n cov\left(\left|U_{k,1}-U_{k,2}\right|,\left|U_{k',1}-U_{k',3}\right|\right)+O(\varepsilon n).\label{eq.covM2}
\end{align}

$M_{3}$. For any realization of $\bm{W}_{k}$ and $\bm{W}_{k'}$, each $i$ shows up twice in the summands of the first summation in $M_{3}$ if $\pi_{k}(i)\neq 1$ and $\pi_{k}(i)\neq n$, and once otherwise. Hence, by i.i.d. and by independence between  $(W_{1},\ldots,W_{K})$ and $(Z_{1},\ldots,Z_{K})$, 
\begin{equation}\label{eq.covM3}
M_{3}=4ncov\left(U_{k,1}(1-U_{k,1}),|U_{k',3}-U_{k',1}|\right)+O(1),
\end{equation}
where $O(1)$ accounts for the case where $\pi_{k}(i)=1$ or $n$.

Similarly,
\begin{equation}\label{eq.covM4}
M_{4}=4ncov\left(U_{k',1}(1-U_{k',1}),|U_{k,2}-U_{k,1}|\right)+O(1).
\end{equation}

Finally,
\begin{equation}\label{eq.covM5}
M_{5}=4ncov\left(U_{k,1}(1-U_{k,1}),U_{k',1}(1-U_{k',1})\right).
\end{equation}

By i.i.d., $(U_{k,1},U_{k',1})\perp (U_{k,2},U_{k',3})$ and $U_{k,2}\perp U_{k',3}$. Letting $U_{k,1},U_{k',1},U_{k,2},U_{k',3}$ be $U_{1},U_{1}',U_{2},U_{3}$ in Lemma \ref{lem.zero}, we have $|M_{1}+M_{2}+M_{3}+M_{4}+M_{5}|\leq C_{3}n\varepsilon$ for some $C_{3}>0$ by combining \eqref{eq.covM1}, \eqref{eq.covM2}, \eqref{eq.covM3}, \eqref{eq.covM4} and \eqref{eq.covM5}. Note that $C_{3}$ does not depend on $\varepsilon$. Substitute it into equation \eqref{eq.cov5} and combine it with \eqref{eq.cov4} and \eqref{eq.cov4.5}. For any $\varepsilon_{0}>0$, let $\varepsilon\leq \varepsilon_{0}/(2C_{3})$. For sufficiently large $n$ such that $C_{1}\Pr(D_{k,k'}(\varepsilon)=0)\leq \varepsilon_{0}/2$, guaranteed by Lemma \ref{lem.de}, we have
\begin{equation*}
\frac{1}{n}\left|cov\left(\sum_{i=1}^{n-1}A_{k,(i)_{k}},\sum_{i=1}^{n-1}A_{k',(i)_{k}'}\right)\right|\leq \Pr(D_{k,k'}(\varepsilon)=0)C_{1}+C_{3}\varepsilon\leq \varepsilon_{0}.
\end{equation*}
We thus obtain the desired result since $\varepsilon_{0}$ is arbitrary.
\end{proof}

Finally, we cite the following central limit theorem for a local-dependent sequence. To save space, please refer to \cite{chatterjee2008new} for the related definitions of an interaction graphical rule, extension of a graphical rule, and Kantorovich-Wasserstein distance.
\begin{lem}[\cite{chatterjee2008new}, Theorem 2.5, p.1589]\label{lem.clt}
Let $\mathcal{X}$ be a measure space and $f:\mathcal{X}^{n}\to\mathbb{R}$ be a measurable map that admits a symmetric interaction rule $G$. Let $X_{1},X_{2},\ldots,$ be a sequence of i.i.d. $\mathcal{X}$-valued random variables and let $\bm{X}=(X_{1},\ldots,X_{n})$. Let $W=f(\bm{X})$ and $\sigma^{2}=\text{Var}(W)$. Let $\bm{X}'=(X_{1}',\ldots,X_{n}')$ be an independent copy of $\bm{X}$. For each $j$, define 
\[
  \Delta_{j}f(\bm{X})=W-f(X_{1},\ldots,X_{j-1},X_{j}',X_{j+1},\ldots,X_{n})
\]
and let $M=\max_{j}|\Delta_{j}f(\bm{X})|$. Let $G'$ be an arbitrary symmetric extension of $G$ on $\mathcal{X}^{n+4}$ and put
\[
\delta\coloneqq 1+\text{degree of the vertex $1$ in }G'(X_{1},\ldots,X_{n+4}).
\]
We then have
\[
\delta_{W}\leq \frac{Cn^{1/2}}{\sigma^{2}}\mathbb{E}\left(M^{8}\right)^{1/4}\mathbb{E}\left(\delta^{4}\right)^{1/4}+\frac{1}{2\sigma^{3}}\sum_{j=1}^{n}\mathbb{E}|\Delta_{j}f(\bm{X})|^{3},
\]
where $\delta_{W}$ is the Kantorovich-Wasserstein distance between the law of $(W-\mathbb{E}(W))/\sqrt{\text{Var}(W)}$ and the standard normal law and $C$ is a universal constant.
\end{lem}
\begin{proof}[Proof of Theorem \ref{thm.joint1}]
Recall we defined $A_{k,(i)_{k}}\coloneqq \left|U_{k,(i+1)_{k}}-U_{k,(i)_{k}}\right|+2U_{k,(i)_{k}}\left(1-U_{k,(i)_{k}}\right)-2/3$ in Lemma \ref{lem.cov}. By Lemma \ref{lem.hajek}, it is sufficient to show that
\[
\begin{pmatrix}
\frac{\sum_{i=1}^{n-1}A_{1,(i)_{1}}}{\sqrt{n-1}}\\
\cdots\\
\frac{\sum_{i=1}^{n-1}A_{K,(i)_{K}}}{\sqrt{n-1}}
\end{pmatrix}\overset{d}{\to}\mathcal{N}\left(\bm{0},\frac{2}{45}I_{K}\right).
\]
By Assumptions \ref{assu.iid} and \ref{assu.cont}, $U_{k,(1)_{k}},\ldots,U_{k,(n)_{k}}$ are i.i.d. Following the proof of Theorem 2 in \cite{angus1995coupling}, we have $\mathbb{E}A_{k,(i)_{k}}=0$ and $\text{Var}\left(\sum_{i=1}^{n-1}A_{k,(i)_{k}}/\sqrt{n-1}x\right)=2/45+o(1)$ for all $k$. Define two subsets of indices $E_{k,i}\coloneqq \{j>i:W_{k,j}=W_{k,i}\}$ and $S_{k,i}\coloneqq\{j:W_{k,j}>W_{k,i}\text{ and }W_{k,j}\leq W_{k,j'}\text{ for all } W_{k,j'}> W_{k,i}\}$. Define the right-nearest neighbor of $i$ for $k$ as:
\begin{equation}\label{eq.rnn}
N_{k}(i)\coloneqq \begin{cases}
\min E_{k,i},&\text{if } E_{k,i}\neq\emptyset;\\
\min S_{k,i},&\text{if } E_{k,i}=\emptyset\text{ and }S_{k,i}\neq \emptyset;\\
i,&\text{if }E_{k,i}=S_{k,i}=\emptyset.
\end{cases}
\end{equation}
This construction guarantees that $N_{k}(i)$ is well-defined even if there are ties in the data. Consider the case where there are no ties, which has probability one under Assumption \ref{assu.cont}. Then $E_{k,i}$ is empty for all $i$. If $W_{k,i}$ is not the largest element among $W_{k,1},\ldots,W_{k,n}$, then $S_{k,i}$ is a singleton and $N_{k}(i)$ is simply the index at which $W_{k,N_{k}(i)}$ is the smallest element that is greater than $W_{k,i}$. If $W_{k,i}$ is the largest, then $S_{k,i}$ is empty as well and $N_{k}(i)$ is defined as $i$.

Let $\bm{X}_{i}=\left(U_{1,i},W_{1,i},\ldots,U_{K,i},W_{K,i}\right)\in\mathcal{X}$ and $\bm{X}=(\bm{X}_{1},\ldots,\bm{X}_{n})\in\mathcal{X}^{n}$. Let function $f_{k,i}(\bm{X})\coloneqq\left(\left|U_{k,i}-U_{k,N_{k}(i)}\right|+2U_{k,i}\left(1-U_{i}\right)-2/3\right)/\sqrt{n}$. By construction, $f_{k,i}$ only depends on $\bm{X}_{i}$ and $\bm{X}_{N_{k}(i)}$. Let function $f_{k}(\bm{X})=\sum_{i=1}^{n}f_{k,i}(\bm{X}_{})$. Since $U_{k,i}$s are bounded, under continuity in Assumption \ref{assu.cont}, we have 
\begin{equation}\label{eq.ab}
\frac{1}{\sqrt{n-1}}\sum_{i=1}^{n-1}A_{k,(i)_{k}}=f_{k}(\bm{X})+o_{p}(1).
\end{equation}
 Fix arbitrary constants $c_{1},\ldots,c_{K}$. Let $f(\bm{X})=\sum_{k=1}^{K}c_{k}f_{k}(\bm{X})$. By the variance of $\sum_{i=1}^{n-1}A_{k,(i)_{k}}/\sqrt{n-1}$ and by Lemma \ref{lem.cov} and equation \eqref{eq.ab}, we then have
\begin{equation}
  \text{Var}\left(f(\bm{X})\right)=\frac{2}{45}\sum_{k=1}^{K}c_{k}^{2}+o(1).
\end{equation}
By the Cram\'er-Wold theorem, and by $\mathbb{E}(f(\bm{X}))=o(1)$, it is thus sufficient to show the following:
\begin{equation}\label{eq.clt.b}
\frac{45\left(f(\bm{X})-\mathbb{E}(f(\bm{X}))\right)}{2\sum_{k=1}^{K}c_{k}^{2}}\overset{d}{\to}\mathcal{N}(0,1)
\end{equation}

We now invoke Lemma \ref{lem.clt} to prove equation \eqref{eq.clt.b} by constructing a symmetric interaction graphical rule $G$ for $f$. The construction is similar to those in the proofs of Theorem 3.4 in \cite{chatterjee2008new} and Lemma 3 in \cite{zhang2023asymptotic}. For each $\bm{x}\in\mathcal{X}^{n}$ and each $k$, define function $d_{k}(i,j;\bm{x})$ on $\{1,\ldots,n\}\times \{1,\ldots,n\}$ as
\begin{equation}\label{eq.indexmapping}
  d_{k}(i,j;\bm{x})=\begin{cases}
   \#\{l:w_{k,i}<w_{k,l}<w_{k,j}\},& \text{if }w_{k,i}\leq w_{k,j};\\
  n+100,& \text{otherwise}.
 \end{cases}
\end{equation}
Equation \eqref{eq.indexmapping} and the definition of $N_{k}(i)$ in equation \eqref{eq.rnn} imply that $d_{k}(i,j;\bm{x})=0$ if $j=N_{k}(i)$.

Given any $\bm{x}$, let $G(\bm{x})$ be the graph on $\{1,\ldots,n\}$ that puts an edge between $i$ and $j$ if and only if there exists an $l$ such that $d_{k}(l,i;\bm{x})\leq 2$ and $d_{k}(l,j;\bm{x})\leq 2$ for at least one $k$. This graphical rule is symmetric because it is invariant to the relabeling of indices. 

Before moving on, let us introduce some notation. Fix $i,j\in\{1,\ldots,n\}$ where $i\neq j$. Given an arbitrary $\bm{x}\in\mathcal{X}^{n}$, let $\bm{x}^{i}$ be formed by replacing $\bm{x}_{i}$ in $\bm{x}$ with a different value $\bm{x}_{i}'$, $\bm{x}^{j}$ be formed by replacing $\bm{x}_{j}$ with $\bm{x}_{j}'$, and $\bm{x}^{ij}$ be formed by replacing $\bm{x}_{i}$ with $\bm{x}_{i}'$ and $\bm{x}_{j}$ with $\bm{x}_{j}'$. 

Now we show that $G$ is an interaction rule for $f$. That is, we show that for every $\bm{x},\bm{x}^{i},\bm{x}^{j},\bm{x}^{ij}\in\mathcal{X}^{n}$ such that $\{i,j\}$ is not an edge in $G(\bm{x})$, $G(\bm{x}^{i})$, $G(\bm{x}^{j})$ and $G(\bm{x}^{ij})$, we have
\begin{equation*}
  f(\bm{x})-f(\bm{x}^{i})-f(\bm{x}^{j})+f(\bm{x}^{ij})=0.
\end{equation*}
By the additive structure of $f$, it suffices to show that for every $k$ and every $l\in\{1,\ldots,n\}$, 
\begin{equation}\label{eq.interactive}
  f_{k,l}(\bm{x})-f_{k,l}(\bm{x}^{i})-f_{k,l}(\bm{x}^{j})+f_{k,l}(\bm{x}^{ij})=0.
\end{equation}
A useful observation is that if $\bm{x},\bm{x}'\in\mathcal{X}^{n}$ and $l,m\in\{1,\ldots,n\}$ are such that $\bm{x}_{l}=\bm{x}_{l}'$ and $\bm{x}_{m}=\bm{x}_{m}'$, then by the definition of $d_{k}$, 
\begin{equation}\label{eq.lip}
  \left|d_{k}(l,m;\bm{x})-d_{k}(l,m;\bm{x}')\right|\leq \#\{r:\bm{x}_{r}\neq \bm{x}_{r}'\}.
\end{equation}
Now we show that equation \eqref{eq.interactive} holds for all $l,k,\bm{x}$ and $\bm{x}'$. 

\textbf{Case 1}. Suppose $d_{k}(l,j;\bm{x})\leq 1$. This immediately implies that $l\neq i$ because otherwise $\{i,j\}$ is an edge in $G(\bm{x})$, a contradiction. 

\textit{Step 1.1.} We first show that 
\begin{equation}\label{eq.xxi}
  f_{k,l}(\bm{x})=f_{k,l}(\bm{x}^{i})
\end{equation}
holds for all $l$ falling into this case and all $k$.

Since $\{i,j\}$ is not an edge in $G(\bm{x})$, under $d_{k}(l,j;\bm{x})\leq 1$, it has to be the case that 
\begin{equation}\label{eq.li}
d_{k}(l,i;\bm{x})> 2
\end{equation}
by definition. Meanwhile, by equation \eqref{eq.lip}, we have 
\begin{equation}\label{eq.dbound}
d_{k}(l,j;\bm{x}^{i})\leq \#\{r:\bm{x}^{i}_{r}\neq \bm{x}_{r}\}+d_{k}(l,j;\bm{x})\leq2.
\end{equation}
Since $\{i,j\}$ is not an edge in $G(\bm{x}^{i})$, equation \eqref{eq.dbound} implies 
\begin{equation}\label{eq.liprime}
  d_{k}(l,i;\bm{x}^{i})> 2.
\end{equation}
By the definition of $N_{k}$ in equation \eqref{eq.rnn}, if $N_{k}(l)=i$ under $\bm{x}$ or $\bm{x}^{i}$, then it has to be case that $d_{k}(l,i;\bm{x})$ or $d_{k}(l,i;\bm{x}^{i})$ is equal to 0. Equations \eqref{eq.li} and \eqref{eq.liprime} thus imply that $N_{k}(l)\neq i$ under both  $\bm{x}$ and $\bm{x}^{i}$. Since all the elements in $\bm{x}$ and in $\bm{x}^{i}$ are equal except for $\bm{x}_{i}$ and $\bm{x}_{i}^{i}$, we have $\bm{x}^{i}_{l}=\bm{x}_{l}$ by $i\neq l$ and $\bm{x}^{i}_{N_{k}(l)}=\bm{x}_{N_{k}(l)}$. Equation \eqref{eq.xxi} thus follows because $f_{k,l}(\bm{x})$ by definition only depends on $\bm{x}_{l}$ and $\bm{x}_{N_{k}(l)}$.

\textit{Step 1.2}. Now we show that 
\begin{equation}\label{eq.xjxij}
  f_{k,l}(\bm{x}^{j})=f_{k,l}(\bm{x}^{ij})
\end{equation}
holds for all $l$ falling into this case and all $k$.

We first show that $d_{k}(l,i;\bm{x}^{j})> 1$. This holds if $l=j$ because $\{i,j\}$ is not an edge in $G(\bm{x}^{j})$. If $l\neq j$, then $\bm{x}^{j}_{l}=\bm{x}_{l}$ and $\bm{x}^{j}_{i}=\bm{x}_{i}$. So we have
\begin{equation*}
d_{k}(l,i;\bm{x}^{j})\geq d_{k}(l,i;\bm{x})-\#\{r:\bm{x}^{j}_{r}\neq \bm{x}_{r}\}> 1,
\end{equation*}
where the first inequality follows \eqref{eq.lip} and the second inequality is by \eqref{eq.li}. 

Similarly, $d_{k}(l,i;\bm{x}^{ij})> 1$ either when $l=j$ or when $l\neq j$ because then
\begin{equation*}
  d_{k}(l,i;\bm{x}^{ij})\geq d_{k}(l,i;\bm{x}^{i})-\#\{r:\bm{x}^{ij}_{r}\neq \bm{x}^{i}_{r}\}> 1,
\end{equation*}
where the first inequality follows \eqref{eq.lip} and the second inequality is by \eqref{eq.liprime}. 

By $d_{k}(l,i;\bm{x}^{j})>1$ and $d_{k}(l,i;\bm{x}^{ij})>1$, following the same argument as in Step 1.1, we conclude that $N_{k}(l)\neq i$ under both  $\bm{x}^{j}$ and $\bm{x}^{ij}$. Hence, we have $\bm{x}^{j}_{l}=\bm{x}^{ij}_{l}$ by $i\neq l$ even if $l=j$, and $\bm{x}^{j}_{N_{k}(l)}=\bm{x}^{ij}_{N_{k}(l)}$. Equation \eqref{eq.xjxij} is proved. Together with Step 1.1, equation \eqref{eq.interactive} is proved for all $l$ such that $d_{k}(l,j;\bm{x})\leq 1$ and all $k$.

\textbf{Case 2}. $d_{k}(l,j;\bm{x}^{i})\leq 1$ or $d_{k}(l,j;\bm{x}^{j})\leq 1$ or $d_{k}(l,j;\bm{x}^{ij})\leq 1$. Equation \eqref{eq.interactive} holds in these three cases following a similar proof as in Case 1 by symmetry. 

\textbf{Case 3}. $d_{k}(l,j;\bm{x})>1$, $d_{k}(l,j;\bm{x}^{i})>1$, $d_{k}(l,j;\bm{x}^{j})>1$ and $d_{k}(l,j;\bm{x}^{ij})>1$. In this case, $l\neq j$. Meanwhile, $N_{k}(l)\neq j$ under $\bm{x}$, $\bm{x}^{i}$, $\bm{x}^{j}$ and $\bm{x}^{ij}$ because otherwise, at least one of the four $d_{k}$s is 0. Therefore, we have $\bm{x}_{l}=\bm{x}^{j}_{l}$ by $l\neq j$ and $\bm{x}_{N_{k}(l)}=\bm{x}^{j}_{N_{k}(l)}$, as well as $\bm{x}_{l}^{i}=\bm{x}^{ij}_{l}$ by $l\neq j$, and $\bm{x}^{i}_{N_{k}(l)}=\bm{x}^{ij}_{N_{k}(l)}$. So, $f_{k,l}(\bm{x})=f_{k,l}(\bm{x}^{j})$ and $f_{k,l}(\bm{x}^{i})=f_{k,l}(\bm{x}^{ij})$. Equation \eqref{eq.interactive} thus holds.

Combining all these cases, we have proved that $G$ is an interaction rule for $f$ because $\bm{x}$ and $\bm{x}'$ are arbitrarily chosen. 

Next, we construct a symmetric extension of $G$ to $(\mathbb{R}^{2K})^{n+4}$. Define a new graph $G'$ such that for any $\bm{x}\in\mathcal{X}^{n+4}$, there is an edge linking $i$ and $j$ if and only if there exists an $l\in\{1,\ldots,n+4\}$ such that $d_{k}(l,i;\bm{x})\leq 6$ and $d_{k}(l,j;\bm{x})\leq 6$ for at least one $k=1,\ldots,K$. One can see that $G'$ is symmetric because it is again invariant to relabeling the indices. To see that $G'$ is an extension of $G$, let $\{i,j\}$ be an edge in $G(\bm{x}_{1},\ldots,\bm{x}_{n})$. Then there exists a $k$ such that for some $l$, $d_{k}(l,i;\bm{x})\leq 2$ and $d_{k}(l,j;\bm{x})\leq 2$ by definition. Now with 4 more points, there are at most 4 more $w_{k,m}$s falling between $w_{k,l}$ and $w_{k,i}$ as well as between $w_{k,l}$ and $w_{k,j}$. This says that $\{i,j\}$ is still an edge in the new graph $G'(\bm{x}_{1},\ldots,\bm{x}_{n+4})$, showing that $G'$ is an extension of $G$. Now we consider the degree of vertex 1 in $G'(\bm{X}_{1},\ldots,\bm{X}_{n+4})$. By continuity of $W_{k}$s, there are no ties with probability 1. So, with probability 1, for any index $i$, there are at most $14K$ $j\neq i$ such that $\{i,j\}$ is an edge in $G'(\bm{X}_{1},\ldots,\bm{X}_{n+4})$. To see this, $j$ can only possibly form an edge with $i$ if there exists a $k$ such that $W_{k,j}$ is among the 7 left-nearest neighbors or the 7 right-nearest neighbors of $W_{k,i}$ since only then there exists an $l$ such that $d_{k}(l,i;\bm{X})\leq 6$ and $d_{k}(l,j;\bm{X})\leq 6$. Therefore, the degree of any vertex of $G'(\bm{X}_{1},\ldots,\bm{X}_{n+4})$ is bounded by $14K$ almost surely, so $\delta$ in Lemma \ref{lem.clt} is 1+14K.

Finally, since $U_{k,i}$ is bounded, there exists a constant $C$ such that $M$ and $\Delta_{j}f(\bm{X})$ in Lemma \ref{lem.clt} are both upper bounded by $C/\sqrt{n}$. Substituting this upper bound, $\delta$ and the variance $2(\sum_{k}c_{k}^{2})/45$ into $\delta_{W}$ in Lemma \ref{lem.clt}, we have $\delta_{W}\to 0$ and thus equation \eqref{eq.clt.b} holds.
\end{proof}

\subsection{Proof of Theorem \ref{thm.joint2}}
\begin{lem}\label{lem.cov.est}
Under Assumptions \ref{assu.iid} to \ref{assum.expansion}, for all $k,k',k''=1,\ldots,K$ and all $i=1,\ldots,n$, 
\begin{equation*}
  cov\left(\sum_{i=1}^{n-1}\left(\left|U_{k,(i+1)_{k}}-U_{k,(i)_{k}}\right|+2U_{k,(i)_{k}}\left(1-U_{k,(i)_{k}}\right)\right),\omega_{(k',k''),i}\left(\bm{Z}_{i}\right)\widetilde{W}_{k'',i}\right)=0.
\end{equation*}
\end{lem}
\begin{proof}

Let $\widetilde{\bm{W}}\coloneqq (\widetilde{\bm{W}}_{1},\ldots,\widetilde{\bm{W}}_{n})$. Following the proof of Lemma \ref{lem.zero} and Lemma \ref{lem.cov},
\begin{equation}
  \mathbb{E}\left(\left|U_{k,(i+1)_{k}}-U_{k,(i)_{k}}\right|+2U_{k,(i)_{k}}\left(1-U_{k,(i)_{k}}\right)\Big| \widetilde{\bm{W}}\right)=\frac{2}{3}.
\end{equation}
Therefore, for all $k,k',k'',(i)_{k},i$,
\begin{align*}
cov\left[\mathbb{E}\left(\left|U_{k,(i+1)_{k}}-U_{k,(i)_{k}}\right|+2U_{k,(i)_{k}}\left(1-U_{k,(i)_{k}}\right)\Big| \widetilde{\bm{W}}\right),\mathbb{E}\left(\omega_{(k',k''),i}\left(\bm{Z}_{i}\right)\widetilde{W}_{k'',i}\Big|\widetilde{\bm{W}}\right)\right]=0.
\end{align*}
Then, by the law of total covariance, for $i$ such that $\pi_{k}(i)\neq (1)_{k}$ or $(n)_{k}$,
\begin{align*}
  &cov\left(\sum_{i=1}^{n-1}\left|U_{k,(i+1)_{k}}-U_{k,(i)_{k}}\right|+2U_{k,(i)_{k}}\left(1-U_{k,(i)_{k}}\right),\omega_{(k',k''),i}\left(\bm{Z}_{i}\right)\widetilde{W}_{k'',i}\right)\\
  =&\mathbb{E}\left[cov\left(\sum_{i=1}^{n-1}\left|U_{k,(i+1)_{k}}-U_{k,(i)_{k}}\right|+2U_{k,(i)_{k}}\left(1-U_{k,(i)_{k}}\right),\omega_{(k',k''),i}\left(\bm{Z}_{i}\right)\widetilde{W}_{k'',i}\Big| \widetilde{\bm{W}}\right)\right]\\
  =&\mathbb{E}\left[\widetilde{W}_{k'',i}\cdot cov\left(2\left|U_{k,i}-U_{k,j}\right|+2U_{k,i}\left(1-U_{k,i}\right),\omega_{(k',k''),i}\left(\bm{Z}_{i}\right)\Big| \widetilde{\bm{W}}\right)\right]\\
  =&\mathbb{E}\left[\widetilde{W}_{k'',i}\cdot cov\left(2\left|U_{k,i}-U_{k,j}\right|+2U_{k,i}\left(1-U_{k,i}\right),\omega_{(k',k''),i}\left(\bm{Z}_{i}\right)\right)\right]\\
  =&0,
\end{align*}
where $j$ in the second equality is an arbitrary index different from $i$. It holds because for any realization of $\widetilde{\bm{W}}$, $i$ shows up twice in 
$\sum_{i=1}^{n-1}\left|U_{k,(i+1)_{k}}-U_{k,(i)_{k}}\right|$ if $\pi_{k}(i)\neq (1)_{k}$ or $(n)_{k}$. By independence between $(Z_{1},\ldots,Z_{K})$ and  $(W_{1},\ldots,W_{K})$ and by i.i.d., the conditional covariance stays constant for all $j\neq i$. The penultimate equality is due to independence between $\bm{Z}$ and $\bm{W}$. The last equality follows from $\mathbb{E}\left(\widetilde{W}_{k'',i}\right)=0$ for all $k''$ and $i$. If $\pi_{k}(i)= (1)_{k}$ or $(n)_{k}$, then the coefficients on $|U_{k,i}-U_{k,j}|$ on the right-hand side of the second and third equality becomes 1 but the final result does not change.
\end{proof}

\begin{proof}[Proof of Theorem \ref{thm.joint2}]
\sloppy Following the notation in the proof of Theorem \ref{thm.joint1}, let $\bm{X}_{i}=(U_{1,i},W_{1,i},\ldots,U_{K,i},W_{K,i})\in\mathcal{X}$ and $\bm{X}=(\bm{X}_{1},\ldots,\bm{X}_{n})\in\mathcal{X}^{n}$, where recall that $U_{k,i}=F_{Z_{k}}(Z_{k,i})$. Let $g_{k,i}(\bm{X}_{i})=\sum_{k'=1}^{K}\omega_{(k,k'),i}(\bm{Z}_{i})\widetilde{W}_{k',i}/\sqrt{n}$. Let $g_{k}(\bm{X})=\sum_{i=1}^{n}g_{k,i}(\bm{X}_{i})$. For $K$ constants $\tilde{\bm{c}}=(\tilde{c}_{1},\ldots,\tilde{c}_{K})'$, let $g(\bm{X})=\sum_{k=1}^{K}\tilde{c}_{k}g_{k}(\bm{X})$. By definition of $\widetilde{W}_{k,i}$ and by Assumption \ref{assu.cont}-ii), we have $\mathbb{E}(g_{k,i}(\bm{X}_{i}))=0$ for all $k$ and $i$. Therefore, by Assumption \ref{assum.expansion}, Lemma \ref{lem.cov.est} and the Cram\'er-Wold theorem, it is sufficient to show that
\begin{equation}\label{eq.clt.est}
  \frac{f(\bm{X})+g(\bm{X})}{\frac{2}{45}\sum_{k=1}^{K}c_{k}^{2}+\tilde{\bm{c}}'\Sigma\tilde{\bm{c}}}\overset{d}{\to}\mathcal{N}(0,1),
\end{equation}
where $f$ is defined in the proof of Theorem \ref{thm.joint1} and $\Sigma$ is defined in Assumption \ref{assum.expansion}.

Again, we invoke Lemma \ref{lem.clt} to prove equation \eqref{eq.clt.est}. We construct the same graphical rule $G$ as in the proof of Theorem \ref{thm.joint1}. We already show that $G$ is symmetric and is an interaction rule for $f$. Now we show that $G$ is also an interaction rule for $g$. That is, for all $\bm{x},\bm{x}^{i},\bm{x}^{j},\bm{x}^{ij}\in\mathcal{X}$ and $\{i,j\}$ is not an edge in the graphs $G(\bm{x}),G(\bm{x}^{i}),G(\bm{x}^{j}),G(\bm{x}^{ij})$, we have
\begin{equation}\label{eq.interactive.est}
  g(\bm{x})-g(\bm{x}^{i})-g(\bm{x}^{j})+g(\bm{x}^{ij})=0.
\end{equation}
By construction, since $g_{k,i}$ only depends on $\bm{X}_{i}$, we have
\begin{align*}
&g(\bm{x})-g(\bm{x}^{i})-g(\bm{x}^{j})+g(\bm{x}^{ij})\\
=&\sum_{k=1}^{K}\tilde{c}_{k}\Big(g_{k,i}(\bm{x}_{i})+g_{k,j}(\bm{x}_{j})-g_{k,i}(\bm{x}_{i}')-g_{k,j}(\bm{x}_{j})-g_{k,i}(\bm{x}_{i})-g_{k,j}(\bm{x}'_{j})+g_{k,i}(\bm{x}_{i}')+g_{k,j}(\bm{x}'_{j})\Big)\\
=&0.
\end{align*}
Therefore, $G$ is an interaction rule for $f+g$. The same extension $G'$ as in the proof of Theorem \ref{thm.joint1} can be constructed as well, so $\delta=14K+1$.

Finally, recall that $|\Delta_{j}f(\bm{X})|$ is upper bounded by $C/\sqrt{n}$ for some $C>0$. Now for $\Delta_{j}g(\bm{X})$:
\begin{align*}
  \left|\Delta_{j}g(\bm{X})\right|\leq& \sum_{k=1}^{K}|\tilde{c}_{k}|\left|g_{k,j}\left(\bm{X}_{j}\right)-g_{k,j}\left(\bm{X}_{j}'\right)\right|\\
  \leq &\frac{1}{\sqrt{n}}\sum_{k=1}^{K}\sum_{k'=1}^{K}|\tilde{c}_{k}|\left|\omega_{(k,k'),j}(\bm{Z}_{j})\widetilde{W}_{k',j}\right|+\frac{1}{\sqrt{n}}\sum_{k=1}^{K}\sum_{k'=1}^{K}|\tilde{c}_{k}|\left|\omega_{(k,k'),j}(\bm{Z}_{j}')\widetilde{W}_{k',j}'\right|.
\end{align*}
Therefore,
\begin{align*}
  \sum_{j=1}^{n}\mathbb{E}\left(\left|\Delta_{j}g(\bm{X})\right|^{3}\right)\leq&\frac{1}{n^{3/2}}\sum_{j=1}^{n}\left[2\sum_{k,k'}|\tilde{c}_{k}|\left[\mathbb{E}\left|\omega_{(k,k'),i}(\bm{Z}_{i})\widetilde{W}_{k',i}\right|^{3}\right]^{1/3}\right]^{3}\\
  \leq &\frac{8K^{6}\max_{k}|\tilde{c}_{k}|^{3}}{\sqrt{n}}\max_{k,k'}\mathbb{E}\left|\omega_{(k,k'),i}(\bm{Z}_{i})\widetilde{W}_{k',i}\right|^{3}\\
  \leq &\frac{8K^{6}\max_{k}|\tilde{c}_{k}|^{3}}{\sqrt{n}}\max_{k,k'}\left[\mathbb{E}\left(\omega_{(k,k'),i}(\bm{Z}_{i})\widetilde{W}_{k',i}\right)^{8}\right]^{3/8},
\end{align*}
where the first inequality is by Minkowski's inequality, the second is by i.i.d., and the last is by Lyapunov's inequality.

For $\left(\max_{j}|\Delta_{j}g(\bm{X})|\right)^{8}$, note that it is equal to $\max_{j}|\Delta_{j}g(\bm{X})|^{8}$. Then,
\begin{align*}
&\mathbb{E}\left(\max_{j}|\Delta_{j}g(\bm{X})|^{8}\right)\\
\leq &\frac{n}{n^{4}}\max_{j}\mathbb{E}\left(\sum_{k,k'}|\tilde{c}_{k}|\left|\omega_{(k,k'),j}(\bm{Z}_{j})\widetilde{W}_{k',j}\right|+\sum_{k,k'}|\tilde{c}_{k}|\left|\omega_{(k,k'),j}(\bm{Z}_{j}')\widetilde{W}'_{k',j}\right|\right)^{8}\\
\leq &\frac{256K^{16}\max_{k}|\tilde{c}_{k}|^{8}}{n^{3}}\max_{k,k'}\mathbb{E}\left(\omega_{(k,k'),j}(\bm{Z}_{j})\widetilde{W}_{k',j}\right)^{8},
\end{align*}
where the last inequality is again by Minkowski's equality. Therefore, $\sqrt{n}\left[\mathbb{E}\left(\max_{j}|\Delta_{j}g(\bm{X})|^{8}\right)\right]^{1/4}=O(1/n^{1/4})$. Substituting these bounds into $\delta_{W}$ in Lemma \ref{lem.clt}, equation \eqref{eq.clt.est} is proved since $\delta_{W}\to 0$.
\end{proof}

\subsection{Proofs of Corollaries \ref{cor.cs} and \ref{cor.csinter}}
To save space, we only prove Corollary \ref{cor.csinter} since the proof of Corollary \ref{cor.cs} follows a similar argument.
\begin{proof}[Proof of Corollary \ref{cor.csinter}]
By equations \eqref{eq.CSindiinter} and \eqref{eq.CSinter}, the event $\bm{\theta}\in CS(1-\alpha)$ is equivalent to the following:
\begin{align*}
&\Pr\left(\bm{\theta}\in CS(1-\alpha)\right)\\
=&\Pr\left(\hat{\theta}_{k}-\frac{c_{k;\alpha}}{\sqrt{n}}\leq \theta_{k}\leq \hat{\theta}_{k}+\frac{c_{k;\alpha}}{\sqrt{n}}\text{ and }\sqrt{n}\xi_{n}\left(P_{k}-\frac{\theta_{k}}{Y_{k}},Z_{k}\right)\leq \sqrt{0.4}z_{(1-\alpha)^{\frac{1}{2K}}},\forall k \right)\\
=&\Pr\left(\hat{\theta}_{k}-\frac{c_{k;\alpha}}{\sqrt{n}}\leq \theta_{k}\leq \hat{\theta}_{k}+\frac{c_{k;\alpha}}{\sqrt{n}}\text{ and }\sqrt{n}\xi_{n}\left(W_{k},Z_{k}\right)\leq \sqrt{0.4}z_{(1-\alpha)^{\frac{1}{2K}}},\forall k \right)\\
=&\Pr\left(\hat{\theta}_{k}-\frac{c_{k;\alpha}}{\sqrt{n}}\leq \theta_{k}\leq \hat{\theta}_{k}+\frac{c_{k;\alpha}}{\sqrt{n}},\forall k \right)\cdot\Pr\left(\sqrt{n}\xi_{n}\left(W_{k},Z_{k}\right)\leq \sqrt{0.4}z_{(1-\alpha)^{\frac{1}{2K}}},\forall k \right)+o(1)\\
=&\sqrt{1-\alpha}\cdot \sqrt{1-\alpha}+o(1)\\
=&1-\alpha+o(1),
  \end{align*}  
  where the first equality is by equations \eqref{eq.CSindiinter} and \eqref{eq.CSinter}, the second equality is by the first order condition \eqref{eq.foc}, the third and fourth equality are by Theorem \ref{thm.joint2} and the validity of the estimation-based confidence band for $\bm{\theta}$.
\end{proof}

\bibliographystyle{chicago}

\bibliography{references}

\end{document}